\documentclass[twocolumn]{aastex631}

\newcommand\sex{{\sc SExctractor} }
\newcommand\alf{{\tt ALF} }

\begin{document}

\title{No Evidence of a Dichotomy in the Elliptical Galaxy Population}

\correspondingauthor{Rog\'erio Monteiro-Oliveira}
\email{rogerionline@gmail.com}

\author[0000-0001-6419-8827]{Rog\'erio Monteiro-Oliveira}
\affiliation{Institute of Astronomy and Astrophysics, Academia Sinica,   
Taipei 106319, Taiwan}

\author[0000-0001-7146-4687]{Yen-Ting Lin}
\affiliation{Institute of Astronomy and Astrophysics, Academia Sinica,   
Taipei 106319, Taiwan}

\author{Wei-Huai Chen}
\affiliation{Department of Physics, National Taiwan University, Taipei 10617, Taiwan}
\affiliation{Institute of Astronomy and Astrophysics, Academia Sinica,   Taipei 10617, Taiwan}

\author[0000-0003-2069-9413]{Chen-Yu Chuang}
\affiliation{Institute of Astronomy, National Tsing Hua University, Hsinchu 30013, Taiwan}
\affiliation{Institute of Astronomy and Astrophysics, Academia Sinica,   Taipei 106319, Taiwan}
\affiliation{Department of Astronomy, University of Arizona, Tucson, AZ 85721, USA}

\author[0000-0002-5258-8761]{Abdurro'uf}
\affiliation{Department of Physics and Astronomy, Johns Hopkins University, Baltimore, MD 21210, USA}
\affiliation{Institute of Astronomy and Astrophysics, Academia Sinica,   Taipei 106319, Taiwan}

\author{Po-Feng Wu}
\affiliation{Department of Physics, National Taiwan University, Taipei 106319, Taiwan}
\affiliation{Institute of Astrophysics, National Taiwan University,   Taipei 106319, Taiwan}

\begin{abstract}

The advent of large integral field spectroscopic surveys has found that elliptical galaxies (EGs) can be classified into two classes: the fast rotators (whose kinematics are dominated by rotation) and the slow rotators (which exhibit slow or no rotation pattern).
It is often suggested that while the slow rotators typically have boxy isophotal shapes, have a high $\alpha$-to-iron abundance ratio, and are quite massive, the fast rotators often exhibit the opposite properties (that is, having disky isophotes, lower $\alpha$-to-iron ratio, and of typical masses).  Whether the EGs consist of two distinct populations (i.e., a dichotomy exists), remains an unsolved issue.  To examine the existence of the dichotomy, we used a sample of 1,895 EGs from the SDSS-IV MaNGA survey, and measured robustly the stellar kinematics, isophotal shapes, and [Mg/Fe] ratio.  We confirmed the previous finding that the bulk of the EGs are disky (65~\%) and fast rotators (67~\%), but found no evidence supporting a dichotomy, based on a principal component analysis.  The different classes (boxy/disky and slow/fast rotators) of EGs occupy slightly different loci in the principal component space. This may explain the observed trends that led to the premature support of a dichotomy based on small samples of galaxies. 

\end{abstract}


\section{Introduction} 
\label{sec:intro}

Located on the ``handle'' of the classical Hubble Tuning Fork diagram, elliptical galaxies (EGs) are often described as morphologically featureless galaxies. Their light is smoothly distributed in an apparently elliptical shape with varied ellipticity, showing no signs of structures like arms and/or bars. However, 
the advent of high resolution imaging and kinematics data have revealed that the EGs do not form a homogeneous group: their isophotes exhibit slight deviations from a perfect ellipse \citep{Bender89}. When the excess of light is detected towards the ``corners'', the EGs are classified as boxy-shaped (boxy EGs hereafter). On the other hand, disky ellipticals (disky EGs hereafter) possess a light excess  towards the major and minor axes. Observations using the Hubble Space Telescope (HST) have shown that the surface brightness profiles in the innermost region ($\ll 1$\,kpc)  also vary \citep{Faber97}, with some galaxies exhibiting a ``core'' (hereafter cored EGs) while others showing a power-law like behavior (hereafter PL EGs). Furthermore, kinematical studies have shown that while the majority of EGs are fast rotating, the most massive ones rotate slowly or exhibit no rotation (please see \citealt{Cappellari16} for a  review).

In an effort to find a pattern in the complex zoo of EGs, \cite{Kormendy96} observed that Hubble's Tuning Fork embeds a sequence governed by the contribution of velocity anisotropy in the  kinematics of galaxies. The anisotropy is small in late type spiral galaxies dominated by rotation and increases towards the elliptical branch, where random motions play an increasingly major role (however, for the most massive EGs, the dispersion becomes mostly isotropic). The same view is shared, for example, by \cite{Krajnovic13} and \cite{Fogarty15}.  \cite{Cappellari11} proposed a  comb-shaped scheme to arrange the spiral, lenticular, and EGs (Figure~1 therein) according to the relative contribution of  rotation compared to random motions, based on integral field spectroscopic (IFS) observations. In this scheme, EGs occupy the  comb's handle. The kinematics is not homogeneous along the handle, however, and two types of EGs can be distinguished, one with stellar content exhibiting a fast rotation pattern (hereafter fast rotators, or FRs), in contrast to those showing  little or no rotation (hereafter slow rotators or SRs) located toward the end of the handle. 

Building upon a series of previous studies,
\citep[e.g.,][]{Davies83,Lauer85b,Nieto89,Tremblay96}, \cite{Kormendy96} 
also proposed the existence of a dichotomy among EGs\footnote{Note that this should not be confused with the distinction between elliptical galaxies and lower mass, elliptical-looking galaxies such as spheroidals, as discussed in Section 2.1 of \cite{Kormendy09}.}, by linking the stellar kinematics to photometric observables. On one side, FRs are expected to be correlated with normal and low-luminosity EGs, whose isophotes are better described by disky shapes. In contrast, the most luminous/massive EGs possess boxy isophotes and are either SRs or even show no rotation (e.g., M87). Later, \citet[][see also \citealt{Lauer12,Krajnovic13b}]{Faber97} found that the innermost light profiles can also serve as a proxy for the stellar kinematics, 
where FRs have primarily power-law distribution, while SRs are cored.

\cite{Kormendy09} presented new data to support the dichotomy of EGs, suggesting that boxy/SRs are strong radio and X-ray emitters, whereas disky/FRs do not show 
noticeable nuclear activity \citep[see also][]{Bender89, Pellegrini99,Sarzi13,Zheng23}. 
The velocity maps of SRs/FRs also implies different 3-dimensional shapes, ranging from  oblate (fully anisotropic) to triaxial (approximately isotropic), respectively \citep{Weijmans14,Foster17}. Furthermore, the stellar populations of boxy/SRs appear to have a higher abundance of $\alpha$ elements when compared with the  disky EGs/FRs. 
The $\alpha$-to-iron ratio, commonly encapsulated by the magnesium-to-iron abundance, [Mg/Fe], is known to correlate strongly with the velocity dispersion $\sigma_\star$ of EGs \citep[e.g.,][]{Thomas05,Zhu10,McDermid15,Watson22}. It is worth mentioning that the observed dichotomy in the [Mg/Fe] abundance advocated by \cite{Kormendy09} was based on  a small ($\approx40$) population of EGs in the Virgo cluster (see their Figure~45), so that a large survey would be beneficial to support (or dispute) this conclusion.

The possible existence of only two dissimilar groups of EGs affords an enormous simplification in the understanding of EG formation \citep[e.g.,][]{Naab06,Hopkins09a, Hopkins09b,Bilek23}. A general picture can be found, for example, in \citet[][see also the references therein]{Cappellari16}  and is illustrated in his Figures 29 and 30. Roughly speaking, under the hierarchical  structure buildup characteristic of the cold dark matter (CDM) model, there would be two main evolutionary tracks to be followed by the EGs. SRs are expected to form earlier, 
within dark matter (DM) halos that collapsed earlier (i.e., higher, rarer peaks of the initial density perturbation) that eventually evolve into
groups or galaxy clusters today, with the SR being the central galaxy of the halo. Inside these halos, inefficient cooling and the presence of hot intracluster/intragroup medium prevent active star formation, in addition to the high duty cycle of the active galactic nucleus (AGN) associated with the central galaxy (the SR). Then, the subsequent growth is driven predominantly by 
dynamical friction; massive satellite galaxies, which could be central galaxies themselves before their own dark matter halos were merged with the main halo, would sink to the center and merge with the SR. The multiple merger events that SRs experienced may gradually reduce their angular momentum \citep{Bois11,Naab14}.

The FRs, on the other hand, are presumed to emerge more frequently in low-density environments (or low-mass DM halos), initially in the form of disk galaxies  with a considerable amount of turbulent gas at $z\gtrsim 2$, originating from either accretion processes or gas-rich minor mergers. Eventually, disk instabilities drive the gas into the galaxy center, forming a bulge. At the same time, star formation is quenched within the group/cluster environment, which also manages to sweep the galaxy's disk through tidal effects. These collections of events can turn the galaxy into a passive FR that, due to the high velocity dispersion of a galaxy cluster, will not be able to grow through massive dry mergers as SRs do. It is thus expected that the present-day SRs are more massive when compared to the bulk of FRs.
Although details are deliberately omitted here, the takeaway message is that the different evolutionary tracks can potentially imprint distinct features in the galaxies observed at $z\approx 0$, not only restricted to their kinematics \citep[e.g.,][]{Kang07}.

Although a compelling scenario, the idea that the population of EGs can be divided into only two families, each exhibiting distinct physical properties, is still a matter of intense debate in the literature. With shape measurements (isophotes and global ellipticity) for 847 early-type galaxies (ETGs) from data release 4 (DR4) of Sloan Digital Sky Survey \citep[SDSS;][]{SDSS-DR4}, \cite{Hao06} found that isophotal shapes are not correlated with the galaxy velocity dispersion, but with the absolute Petrosian magnitude  ($M_{\rm r}$) and the ellipiticity (in the sense that the boxy fraction increases with  mass and luminosity). Furthermore, only a mild trend for disky EGs being radio-loud was reported. 
Using the same shape measurements, \cite{Pasquali07} confirmed the tight correlation of the isophotal families with both the magnitude ($M_{\rm B}$) and the galaxy dynamical mass. The disky fraction, for example, can reach values as high as 80\% at $M_{\rm B}\sim -19.5$ and $M_{\rm dyn}\sim 8\times10^{10}$\,M$_\odot$, decreasing to 50\% at the brightest and more massive end, showing a continuous transition between boxy and disk EGs (rather than a dichotomy). Later, \cite{Mitsuda17} found that this fraction keeps unchanged since $z\approx 1$, suggesting no evolution in the past 8\,Gyr. Regarding the connection between light distribution and the AGN activity, \cite{Pasquali07} rejected the hypothesis that isophotal shapes trace the radio emission. This is also in line with \cite{Krajnovic20}, wherein HST observations of the inner parts revealed a considerable number of EGs with seemingly contradictory properties (i.e., PL SR and cored FR). Another strong  argument disfavoring the dichotomy scenario was given by \cite{Emsellem11}, who, taking advantage of  high fidelity IFS data, derived the kinematics classification from the specific angular momentum $\lambda$ and found that  boxiness/diskiness is  not a good proxy for the stellar kinematics, same conclusion reached by \cite{He14}. Finally, according to \cite{Chaware14}, the classification of the isophotal shapes might change from the inner to the outer parts of galaxies, making the classification of their  light distribution ambiguous. As  can be seen, there are convincing arguments suggesting that the classification into only two classes may not be suitable to account for the diversity of EGs.

This intense and (so far) inconclusive debate on the reliability of the dichotomy scenario motivates us to investigate a diverse set of parameters of EGs, including the (global) isophotal light distribution, absolute $r$-band magnitude, stellar mass, $\alpha$-element abundance in terms of [Mg/Fe], stellar angular momentum $\lambda$, ellipticity, and the stellar velocity dispersion $\sigma_\star$, and how they are interconnected. The selection of parameters was guided by those employed in \cite{Kormendy09}, derived either directly or indirectly from ground-based observations. Our goal is to obtain high-quality measurements of these parameters for a large sample of EGs and to critically assess the reality of their proposed dichotomy, thereby testing the scenario outlined in \cite{Kormendy09}. To this end, we resort to the principal component analysis (PCA), a powerful statistical technique designed to reduce the dimensionality of a dataset, making it easier to identify any possible patterns there.

In this work, we analyze a large sample of 1895 galaxies at $z<0.15$ identified as EGs via deep learning from the SDSS-IV MaNGA survey \citep{Bundy15}. From the exquisite data provided by the two-dimensional spectral measurements of MaNGA, we measure the stellar kinematics, and determine the  [Mg/Fe] ratio. To test the link between kinematic and chemical properties with a more direct observable, we measured the shape of each galaxy's two-dimensional (2D) light distribution using imaging data from SDSS. Our final catalog also includes the corresponding absolute magnitudes and stellar mass obtained from the NASA-Sloan Atlas\footnote{https://www.sdss4.org/dr13/manga/manga-target-selection/nsa/}.
This rich set of 13265 measurements\footnote{7 parameters for each of 1895 EGs.} will  shed light on the intense debate on the existence of dichotomy.

This paper is organized as follows: Our sample of EGs is introduced in Section~\ref{sec:data}, and the process of obtaining their physical properties is described in detail in Section~\ref{sec:properties}. The PCA and further investigation of correlations among the galaxies' properties are presented in Section~\ref{sec:exploratory}. We discuss our findings in Section~\ref{sec:discussion} and summarize our conclusions in Section~\ref{sec:summary}.

Throughout the paper, we adopt the standard flat cosmology described by $\Omega_M=0.27$, $\Omega_\Lambda=0.73$,  and $H_0=70$ km~s$^{-1}$~Mpc$^{-1}$. Unless stated otherwise, all quoted error bars refer to 68\% of the confidence level (c.~l.).

\section{Data and Sample Galaxies} 
\label{sec:data}

\subsection{MaNGA} 
\label{sec:data.manga}

We use the IFS data from the Mapping nearby Galaxies at Apache Point Observatory \citep[MaNGA;][]{Bundy15} survey to get the spectra of our EG sample. Specifically, we use the final data release of SDSS-IV (DR17; \citealt{Abdurrouf22}), which includes over 10,000 galaxies at $0.01<z<0.15$. The MaNGA hexagonal fiber bundles make use of the BOSS (Baryonic Oscillation Spectroscopic Survey) spectrographs \citep{Smee13} and observe spectra over a wide wavelength range of $3600-10300$~\AA\ with a spectral resolution of $R\sim 1100-2200$. With dithering, the resulting data cubes have an effective spatial resolution FWHM (full width at half maximum) of $2\farcs5$ and spatial sampling of $0\farcs5$ $\rm{spaxel}^{-1}$ \citep{Law15}. We use the \texttt{LOGCUBE} data cubes from the Data Reduction Pipeline \citep[DRP;][]{Law16} for the analysis in this paper. The data cubes have a typical $10\sigma$ limiting continuum surface brightness of $23.5$ mag arcsec$^{-2}$ over a $5\arcsec$ diameter aperture in the $g$-band \citep{Law16}. To get a representative spectrum of each galaxy in our sample, we integrate spectra over all pixels within the effective radius $R_{50}$ (see Section~\ref{sec:nsa}).

\subsection{SDSS}
\label{sec:data.sdss.imaging}

The imaging data used to characterize the 2D light shape of the EGs (i.e., boxy vs.~disky) is also retrieved from SDSS DR17. We use imaging data in the $r$-band, which correspond to the deepest filter observed ($r_{\rm lim}=22.2$ AB mag; 95\% confidence level~for point sources detection). Aiming to map the  light distribution out to large radii, we obtain relatively wide-field imaging cutouts, namely $800\times800$ pixels (0.396$^{\prime\prime}$/pixel), to ensure that even the largest EGs \citep[typically $R_{50}\sim$100--200 arcsec; e.g.,][]{Kormendy09} will have their outermost isophotes confidently measured.

\subsection{Galaxy Zoo} 
\label{sec:gal.zoo}

The Galaxy Zoo project \citep[GZ1; ][]{Lintott08} started as a community effort to visually classify the morphology of a large sample of galaxies imaged by the SDSS. Recognized as featureless galaxies, i.e., lacking apparent optical structures like arms and/or bars, the identification of EGs is relatively straightforward when compared to other classes.

A step further in identifying a larger variety of galaxy features was introduced by  Galaxy Zoo 2 \citep[GZ2;][]{Willett13} project. 
This time, volunteers re-checked the brightest quarter of the GZ1 sample, but now adopting a more branched decision tree, which allowed the identification of a more diverse population of galaxies. In the  case of EGs, even though being considerably simpler than other types of galaxies, they were also classified according to their roundness level. As GZ2 encompasses 96\% of the DR17 MaNGA galaxies\footnote{https://www.sdss4.org/dr17/data\_access/value-added-catalogs/?vac\_id=galaxy-zoo-classifications-for-manga-galaxies}, we also take into account of their morphological classification in our sample selection described below.

\subsection{Sample Selection} 
\label{sec:data.sample}

We select our sample of local EGs based on the DR17 version of the MaNGA Deep Learning  Morphological catalog \citep{DSanchez22, DSanchez18, Fischer19}. This catalog provides morphological information of 10,127 MaNGA galaxies that are classified using the Convolutional Neural Networks (CNNs) that have been trained on two morphological catalogs based on visual inspection: \citet{Nair10} and Galaxy Zoo 2 \citep{Willett13}. The CNN algorithm takes as input composite RGB images of SDSS DR7 galaxies \citep{SDSS-DR7}. To get a clean sample of EGs, we follow the instruction from \citet{DSanchez22} to set the following requirements: \texttt{T-Type} $\leqslant 0$, \texttt{P\_S0} $\leqslant 0.5$, \texttt{P\_LTG} $<0.5$, and \texttt{VC} $=1$.  \texttt{T-Type} is related to the relative importance of the bulge and disk components, while the \texttt{P\_S0} and \texttt{P\_LTG} are the probability of a galaxy being S0 rather than pure elliptical and the probability of a galaxy having a late-type morphology, respectively. The last parameter, \texttt{VC}, is a flag for which the value is set to 1 when a galaxy is visually confirmed to be an elliptical by \citet{DSanchez22}. This selection criterion is designed to align with the classical EGs, i.e., the left-hand side of the Hubble Tuning Fork, thereby matching a similar sample considered in previous studies on dichotomy \citep[e.g.,][]{Kormendy09, Kormendy12}. With these criteria and considering only galaxies whose properties are successfully measured by our pipeline (see Section~\ref{sec:properties}) we end up  with 1895 EGs, which we will refer to  as the MaNGA sample.

\begin{figure*}[ht!]
\begin{center}
\includegraphics[width=\textwidth]{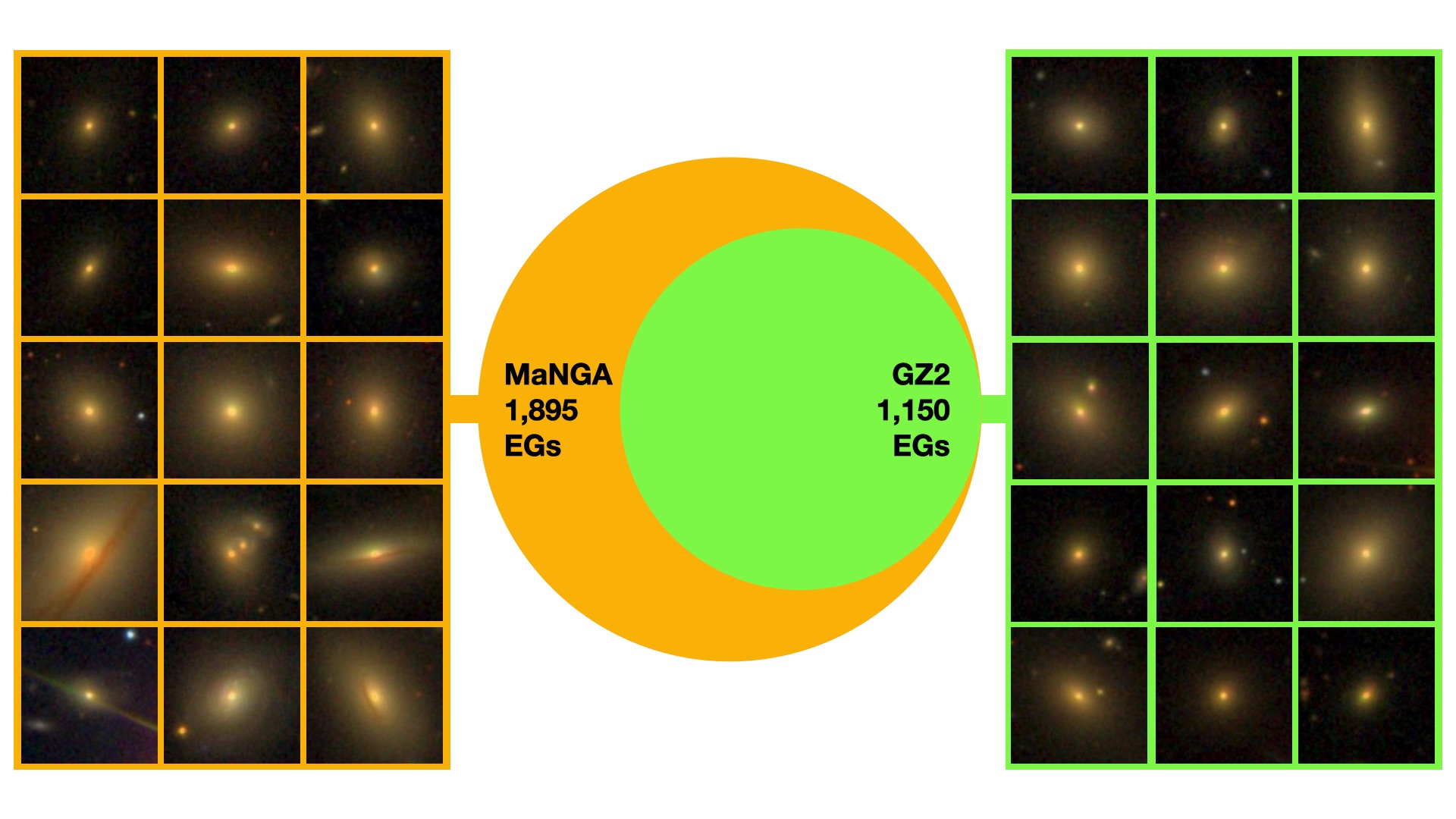} \caption{Cutouts of some EGs in this work taken from SDSS DR17. The MaNGA sample (orange) is comprised by 1895 EGs, of which 1150 overlap with those classified as featureless galaxies by the Galaxy Zoo 2 project (green).}
\label{fig:sample}
\end{center}
\end{figure*}

Although machine learning techniques have a growing role in analyzing astrophysical data, boosted by the unprecedented amount of information provided by current and scheduled surveys, factors like the object's small size, viewing angle, or even the presence of a companion could lead to the misclassification of a non-negligible number of galaxies. Thus, aiming to increase the purity of the EGs to be analyzed in this work, we cross-match the MaNGA sample with the GZ2 galaxies classified as featureless by the volunteers. We identified 1150 galaxies in common, which we will refer to as the GZ2 sample. In Figure~\ref{fig:sample}, we show some examples of EGs in the MaNGA and GZ2 samples.

\section{Properties of our Sample Galaxies} 
\label{sec:properties}

\subsection{NASA-Sloan Atlas} 
\label{sec:nsa}

\begin{figure}
\begin{center}
\includegraphics[width=\columnwidth]{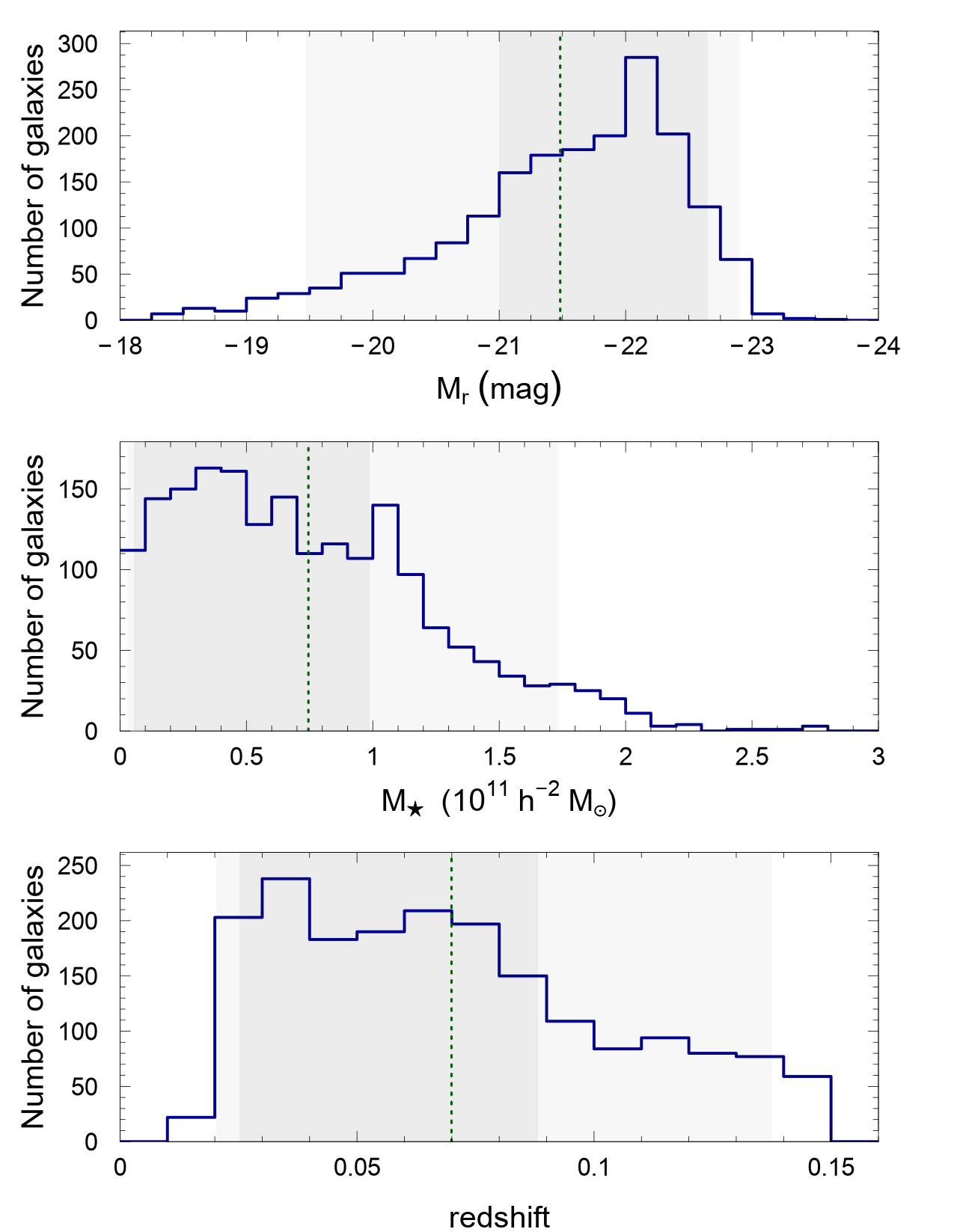} \caption{
Absolute $r$-band magnitude (top), stellar mass (center), and redshift (bottom) distribution for the galaxies in our  sample. The dark (light) shaded area corresponds to the  68 (95) percentile range, and the dotted line represents the corresponding mean value.}
\label{fig:basic}
\end{center}
\end{figure}

To obtain accurated photometry, we use the NASA-Sloan Atlas (NSA; \texttt{v1\_0\_1}) produced by M.~Blanton \citep[e.g.,][]{Blanton11}.  Specifically, we use the elliptical Petrosian photometry from NSA, such as half-light (or effective) radius $R_{50}$ in the $r$-band (\texttt{ELPETRO\_TH50\_R}), $R_{90}$ (\texttt{ELPETRO\_TH90\_R}), which encloses 90\% of the total light, as well as the absolute magnitude, \texttt{ELPETRO\_ABSMAG} ($M_{\rm r}$), the stellar mass from the \texttt{K-correct} fit, \texttt{ELPETRO\_MASS} ($M_{\star}$, in units of $h^{-2}$\,M$_\odot$), and the spectroscopic redshift $z$. A summary showing the distribution of our sample in $M_r$, $M_\star$, and $z$  is presented in Figure~\ref{fig:basic}.

\subsection{From Spectrum to Physical and Chemical Parameters} 
\label{sec:chemical}

To extract physical properties (e.g., stellar velocity dispersion, $\sigma_\star$) and chemical abundances from the MaNGA-based spectra (Section~\ref{sec:data.manga}), we resort to the software package Absorption Line Fitter ({\tt ALF}\footnote{https://github.com/cconroy20/alf}; \citealt{Conroy12,Conroy14,Conroy18}). The code works by modelling an old ($\gtrsim 1$ Gyr) stellar population, allowing for a large range of metallicities [$-2.0:+0.25$] and extending from the optical (0.35 $\mu$m) to the near-infrared (2.4 $\mu$m). \alf embodies the Modules for Isochrones and Stellar Tracks \citep[MIST;][]{Choi16} and the MILES empirical stellar library \citep{Sanchez-Blazquez06,Villaume17}. The initial mass function is set to the \cite{Kroupa01} form in this work.

The parameter space is sampled by the Monte Carlo Markov Chains (MCMC) algorithm {\tt emcee} \citep[][]{Foreman-Mackey13} following the number of walkers ($N_{\rm w}$), burn-in interactions ($N_{\rm b}$), and steps ($N_{\rm s}$) input by the user. These numbers should provide a better compromise between the processing time, precision, and accuracy of the final posteriors. Given the low redshift of our sample (mean redshift $\bar{z}=0.07$, see Figure~\ref{fig:basic}), we optimize the code execution by restricting the wavelength domain to the interval 3800--5800~\AA. \alf has two modes. The so-called ``simple mode'' returns 13 parameters, including those of interest to our study, $\sigma_\star$ and the abundances of iron [Fe/H] and magnesium [Mg/H]. The desired abundance can be obtained as
\begin{equation}
    [{\rm Mg/Fe}] = [{\rm Mg/H}] - [{\rm Fe/H}] \,.
    \label{eq:abundance}
\end{equation}

On the other hand, the ``full mode'' adds 29 more parameters\footnote{A complete description can be found in \cite{Conroy18}.}, which  increase the modeling complexity and result in dramatically longer run time. Following a performance test with 20 spectra described in Appendix~\ref{sec:alf.bench}), we run \alf in the simple mode.

\begin{figure*}[ht!]
\begin{center}
\includegraphics[width=\textwidth]{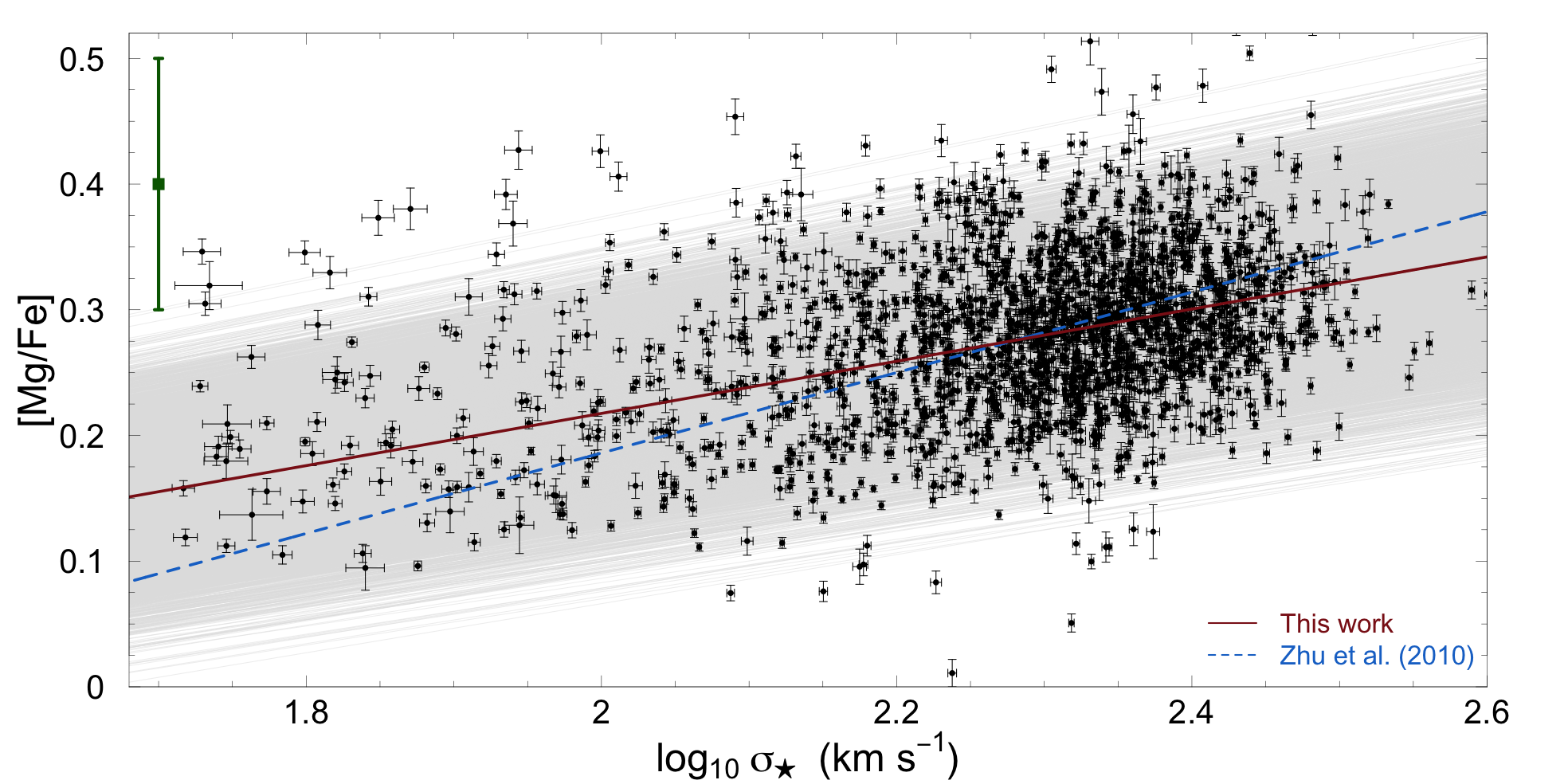} \caption{Magnesium-to-iron ratio [Mg/Fe] and stellar velocity dispersion $\sigma_\star$ of the MaNGA sample extracted by {\sc alf}. The best coefficients of the linear regression modeled by the maximum likelihood estimation are represented by the red straight line. The uncertainties in the fit are illustrated by $10^4$ lines randomly drawn from the posterior distribution (shown in gray). For comparison, we overlaid the best-fit found by \cite{Zhu10}, which is fully consistent with our dataset within the error bars. The error bar in the upper-left hand side illustrates the median error bar in [Mg/Fe] measured from SDSS spectra in to \cite{Zhu10}.}
\label{fig:alf.output}
\end{center}
\end{figure*}

Some of the outputs from \alf relevant to this work is shown in Figure~\ref{fig:alf.output}. To prevent the final catalog from having unreliable measures, we remove  nonphysical values of $\sigma_\star$ by considering only galaxies within the interval
$50 \ {\rm km \ s}^{-1} \leq \sigma_\star \leq 500 \ {\rm km \ s}^{-1}$. Additionally, we discard any measurement with unusually large errors  (e.g., larger than the fraction encompassing 95\% of the sample). This procedure resulted in 162 galaxies being discarded.

The Pearson's correlation coefficient of $0.38\pm0.02$ points to a moderate degree of correlation between [Mg/Fe] and $\sigma_\star$. Thus, we fit the data with a linear regression modeled by the maximum likelihood estimation 
(MLE; using the {\sc R} package {\tt deming}; \citealt{deming}).
The red line in Figure~\ref{fig:alf.output} shows the best values of the intercept and slope of the best-fit, respectively: $-0.20\pm0.03$ and $0.209\pm0.014$. To demonstrate the  uncertainty of the fit, $10^4$ gray lines randomly drawn from the coefficients' probability distribution were overlaid. As a comparison, we show the fit based on 1923 elliptical galaxies at $z < 0.05$ presented by \cite{Zhu10}. Our linear regression is fully consistent with their analysis, but with an important advantage, since our abundances  extracted from the high-quality MaNGA spectra have considerable smaller error bars when compared to those from the SDSS spectra (one tenth on average; see the representative error bar in Figure~\ref{fig:alf.output} and more details in Appendix~\ref{sec:alf.bench}) and also cover a wider range of redshifts.  The complete catalog of [Mg/Fe] and $\sigma_\star$ obtained in this work is presented in Appendix~\ref{sec:parameters}.

\subsection{Shape of the Light Distribution} 
\label{sec:light}

Isophotes represent the edges of uniform surface brightness and typically deviate from a perfect elliptical shape for elliptical galaxies. An excess of light may be observed either along the major axis, leading to a disky shape, or at the ``corners'' of the base ellipse, resulting in a boxy shape. Resorting to SDSS images (Section~\ref{sec:data.sdss.imaging}),  we measured the diskiness/boxiness, i.e., the difference between the real isophote and the best-fit ellipse of our MaNGA sample.

\begin{figure*}
\begin{center}
\includegraphics[width=\textwidth]{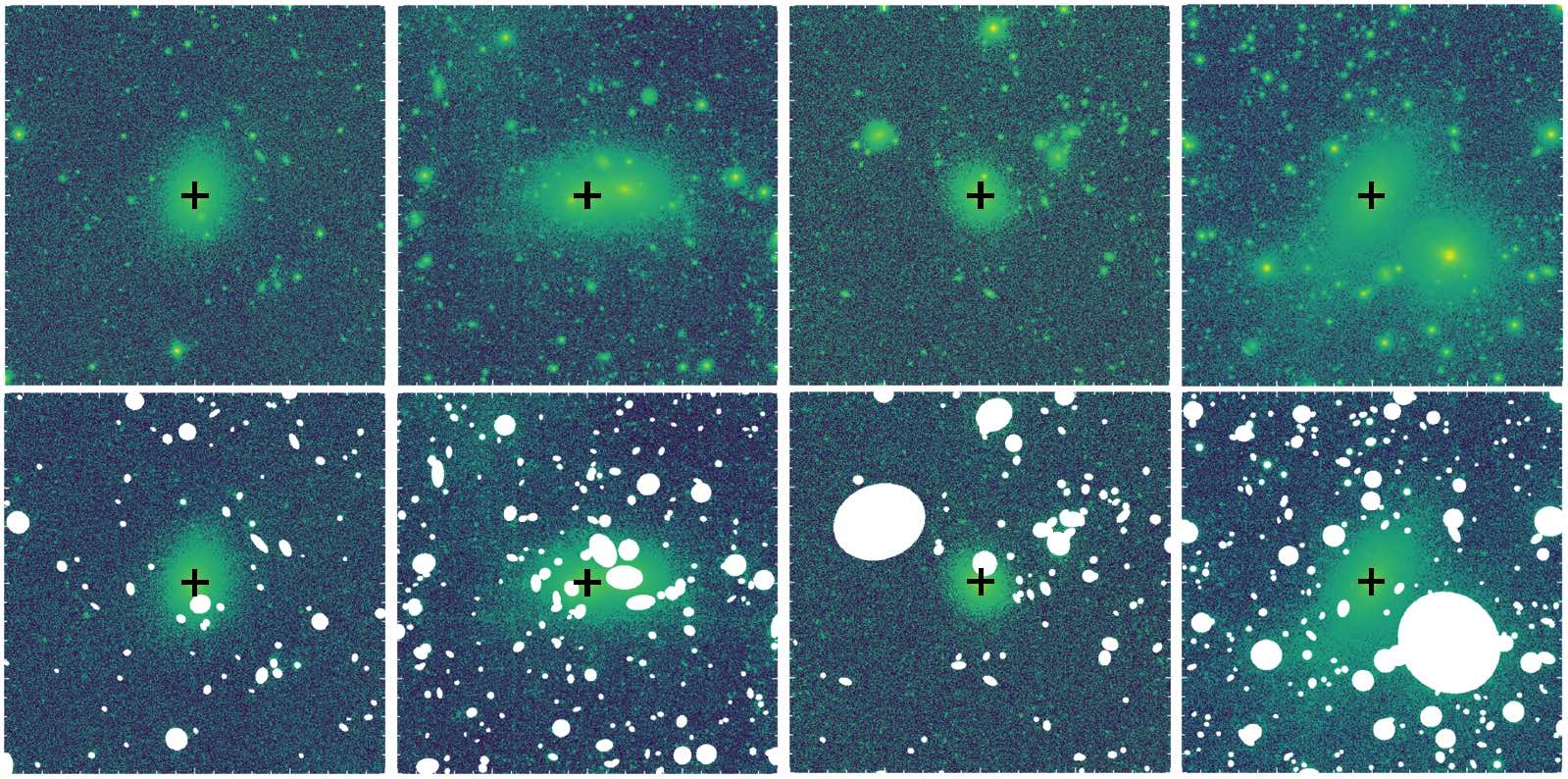} \caption{Examples of the masking procedure applied to the images of four galaxies in our sample. The first row refers to the original images, while the second shows the applied masks. The symbol `+' indicates the center of the galaxy, which was kept out of any mask.}
\label{fig:mask}
\end{center}
\end{figure*}

Prior to fitting the light distribution, we first masked all objects other than the target galaxy. We ran \sex \citep[][]{sextractor} to identify the stars and galaxies in the field. In order to mask the bright, saturated stars and aim to preserve as much of the background area as possible, we defined different relative sizes for the elliptical masks. To this end, we considered both the Kron radius (which encompasses 90\% of the flux) and the {\tt CLAS\_STAR} parameter. The star-like objects ({\tt CLAS\_STAR}$>0.2$) were masked by an ellipse with $3.0\times${\tt \{A,B\}\_IMAGE} ($A$ and $B$ are the ellipse's major and minor axes, respectively.), large enough to encompass bright stars, as determined through empirical observation. The same setup was applied to galaxy-like objects ({\tt CLAS\_STAR}$\leq0.2$) but with a {\tt FLUX\_AUTO} that exceeded the corresponding 95 percentile of the sample. For the other galaxies, a factor of $1.1\times${\tt \{A,B\}\_IMAGE} proved to be enough. We present a few examples of masked fields in Fig.~\ref{fig:mask}. The ``cleaned'' image comprises only the EG plus the background. To avoid any miscentering in the forthcoming modeling of the light distribution, we ensured that the center of the EG was not masked. If that happened, we reduced the size of the mask until the center was clear.

The masked images also enabled us to check the quality of the sky subtraction. For this purpose, we removed the central EG by masking it according to the more restrictive criteria described above. We found a background level comparable to zero with a tiny standard deviation, $0.003\pm0.23$ counts.

We used the {\sc iraf} package {\tt ellipse} \citep{Jedrzejewski87} to decompose the EG isophotal light distribution as a Fourier series
\begin{equation}
    I(\theta)=I_0+\sum_{n=1}^N[A_n \sin(n\theta)+B_n\cos(n\theta)]
    \label{eq.I.theta}
\end{equation}
fitted up to the $N$-th highest order (in practical, $N\leq 4$) and considering the azimuthal angle $\theta$ taken from the major axis. The coefficients of the Fourier harmonics $A_n$ and $B_n$ are directly related to physical shape parameters like the ellipse's center ($n=1$), the position angle and ellipticity ($n=2$), and different aspects of asymmetries ($n=3,4$), to quote only the smallest orders \citep[see e.g. Fig.~1 in][]{Ciambur15}. In particular, $B_4$ (or generally speaking, the fourth cosine coefficient) describes the diskiness/boxiness level as it computes the deviation from a perfect ellipse along the major axis of each isophote.

In order to allow a direct comparison with previous works \citep[e.g.,][]{Bender88, Hao06,Chaware14}, we adopted here the structural parameter $a_4/a$ as defined by \cite{Bender88}. Although the {\tt ellipse} output $B_4$ is normalized by the semi-major axis $a$ and the local gradient intensity ($dI/da$), we can re-scale it in terms of the equivalent radius $r=\sqrt{ab}$ \citep[][where $a$ and $b$ are the major and minor axes of an ellipse, respectively]{Milvang-Jensen99} and obtain $a_4/a$ as follows:
\begin{equation}
    \frac{a_4}{a}=B_4 r/a=B_4\sqrt{1-\epsilon},
    \label{eq.a4}
\end{equation}
where  the ellipticity \begin{equation}
\epsilon=1-b/a 
\label{eq.epsilon}
\end{equation}
is computed from the isophotal's minor and major axes. An isophote with a boxy shape is expected to have $a_4/a<0$, while $a_4/a>0$ describes a disky-like appearance. We fed {\tt ellipse} with initial guesses provided by the \sex catalog (center coordinates, ellipticity, and position angle). In our model, we allowed these quantities to vary freely.

\begin{figure*}
\begin{center}
\includegraphics[width=0.9\textwidth]{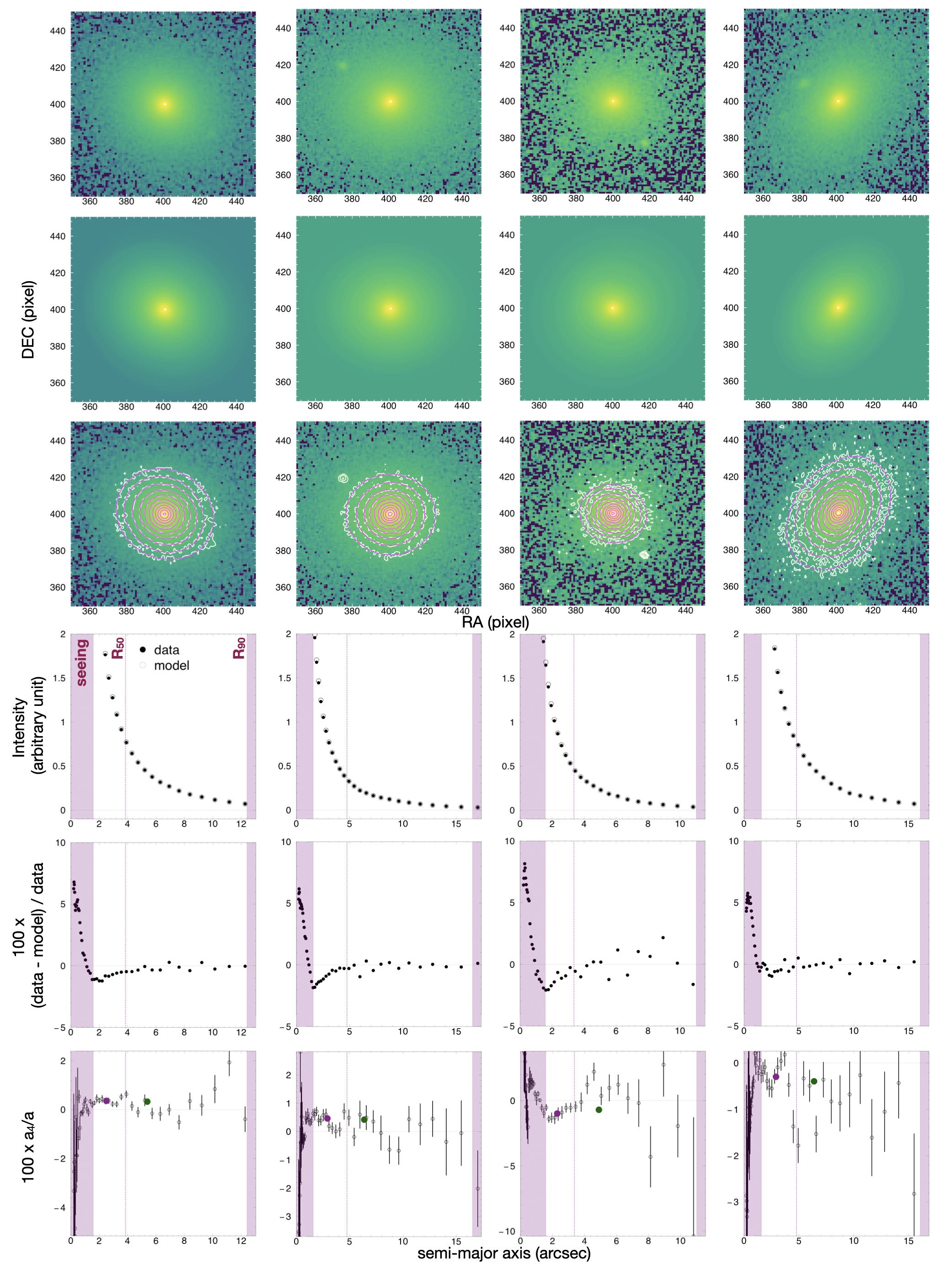} \caption{Process of obtaining the light distribution shape illustrated for four galaxies (one per column). Starting from the original galaxy image (top row), we modeled its light distribution using {\tt ellipse} (second row) and determined their isophotes (third row). These are the central cuts of the original SDSS images, shown in the logarithmic scale. Overall, the model shows a good agreement with the data, as attested by the 1D-light distribution profile (fourth row) and the corresponding relative residuals (fifth row). We have excluded from the analysis the isophotes within both the inner core affected by the seeing (1.58 arcsec) and the outer regions beyond $R_{90}$. Finally, the weighted shape parameter $ \langle a_4/a \rangle$ was obtained in two different regions, inside the effective radius $R_{50}$ (gray vertical line) and $R_{90}$. We only considered in our final sample galaxies for which diskiness/boxiness estimates are consistent within the two radii.}
\label{fig:ellipse.summary}
\end{center}
\end{figure*}

Take $a_4/a$ individually for each isophote (e.g., as a function of the radius) is not meaningful for two main reasons, as (1) we aim to correlate the shape with global galaxy properties, and (2) a bad fit of the light distribution at a given radius can potentially lead to changes in the shape classification \citep[e.g.,][]{Kormendy09}. Instead, we computed the weighted mean $a_4/a$ \citep[e.g.,][]{Chaware14}
\begin{equation}
  \left \langle  \frac{a_4}{a} \right \rangle =\frac{ \int_{r_s}^{R^\prime} \frac{a_4}{a} (r) I(r)[\sigma_{a_4/a}(r)]^{-2}\, dr}
  { \int_{r_s}^{R^\prime} I(r)[\sigma_{a_4/a}(r)]^{-2}\, dr}
    \label{eq.a4.pond}
\end{equation}
where $\sigma_{a_4/a}$ represents the standard deviation of $a_4/a$ over the integration interval, with the lower limit corresponding to the size of the seeing disk, 
$r_s=1.58$~arcsec\footnote{lower quartile; \\ http://www.sdss2.org/dr7/products/general/seeing.html}. To check the consistency of the shape measurements in different regions, we set $R^\prime=R_{50}$ and $R^\prime=R_{90}$. For both cases, the final error on $ \langle a_4/a \rangle$ was obtained via bootstrapping. The process of measuring the light distribution  is illustrated in Fig.~\ref{fig:ellipse.summary}.

In general, the uncertainties on $a_4/a$ scale  with the galactocentric distance, but as Eq.~\ref{eq.a4.pond} down weighs  those data points, they do not have a strong impact on $\langle a_4/a \rangle$. The {\tt ellipse} models provided a good match with the observational data. This is corroborated by the low relative residual in each EG, around 5\% even at larger radii. As a consequence of these high-quality light profiles, we obtained shape measures consistent between the different values of integration boundaries ($R'$). Of the 2047 galaxies we started with (note this value is somewhat larger than those introduced in Section~\ref{sec:data.sample}), only 8\% of the galaxies showed a divergent classification between $R_{50}$ and $R_{90}$. Excluding them, we ended up with 1895 EGs, as mentioned in Section~\ref{sec:data.sample}. Their shape-based classification can be seen in Fig.~\ref{fig:ellipse}. The majority, 1232 EGs (65\%), were classified as disky-shaped. In Appendix~\ref{sec:ellipse.test}, we provide a detailed benchmark test of the {\tt ellipse} package to evaluate its accuracy in recovering elliptical-shape deviations. The measured face values of $\langle a_4/a \rangle$ from this study can be found in Appendix~\ref{sec:parameters}.

\begin{figure}
\begin{center}
\includegraphics[width=\columnwidth]{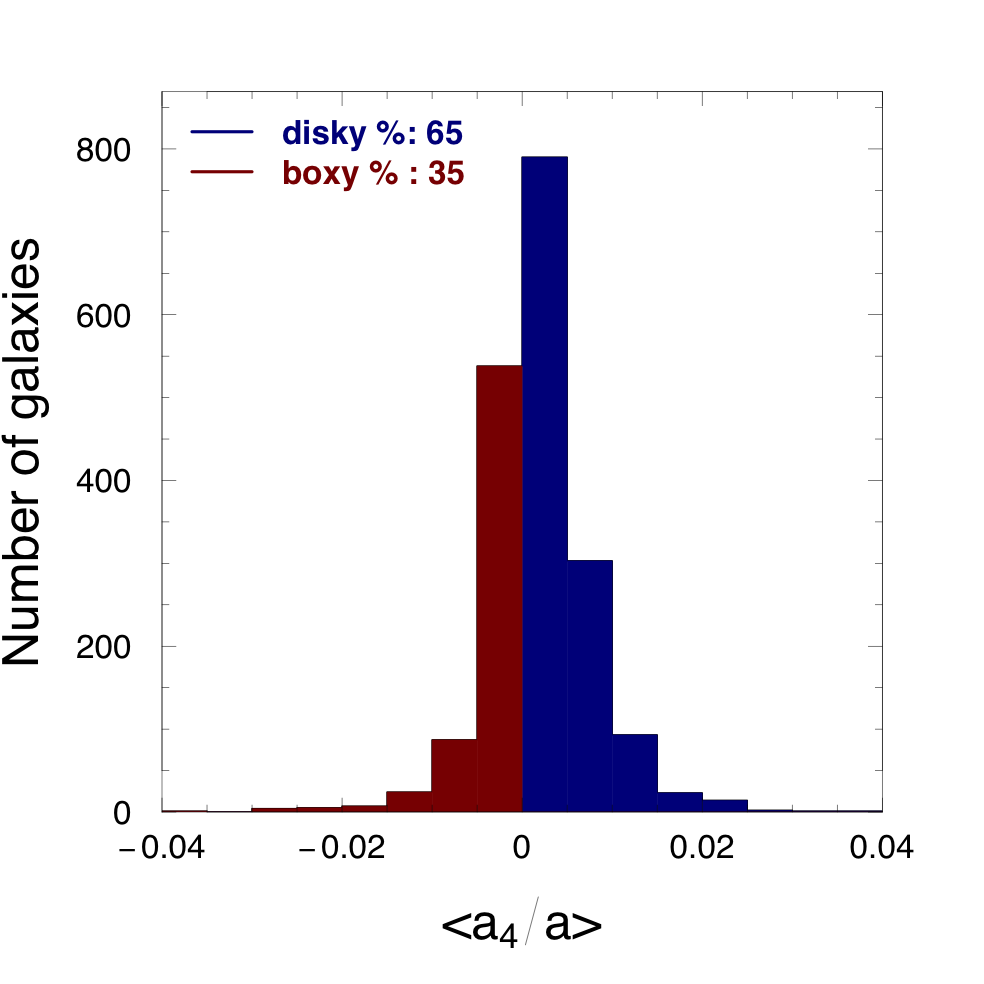}  \caption{Classification of the light distribution shape of our MaNGA sample. EGs are called disky-shaped if $\langle a_4/a \rangle >0$ (Eq.~\ref{eq.a4.pond}), whereas those with a boxy-like appearance have $\langle a_4/a\rangle>0 $. To construct our final sample, we selected only galaxies with consistent measurements of $\langle  a_4/a \rangle $ inside both $R_{50}$ and $R_{90}$ radii.}
\label{fig:ellipse}
\end{center}
\end{figure}

\subsection{Stellar Kinematics} 
\label{sec:dynamics}

\begin{figure*}
\begin{center}
\includegraphics[width=\textwidth]{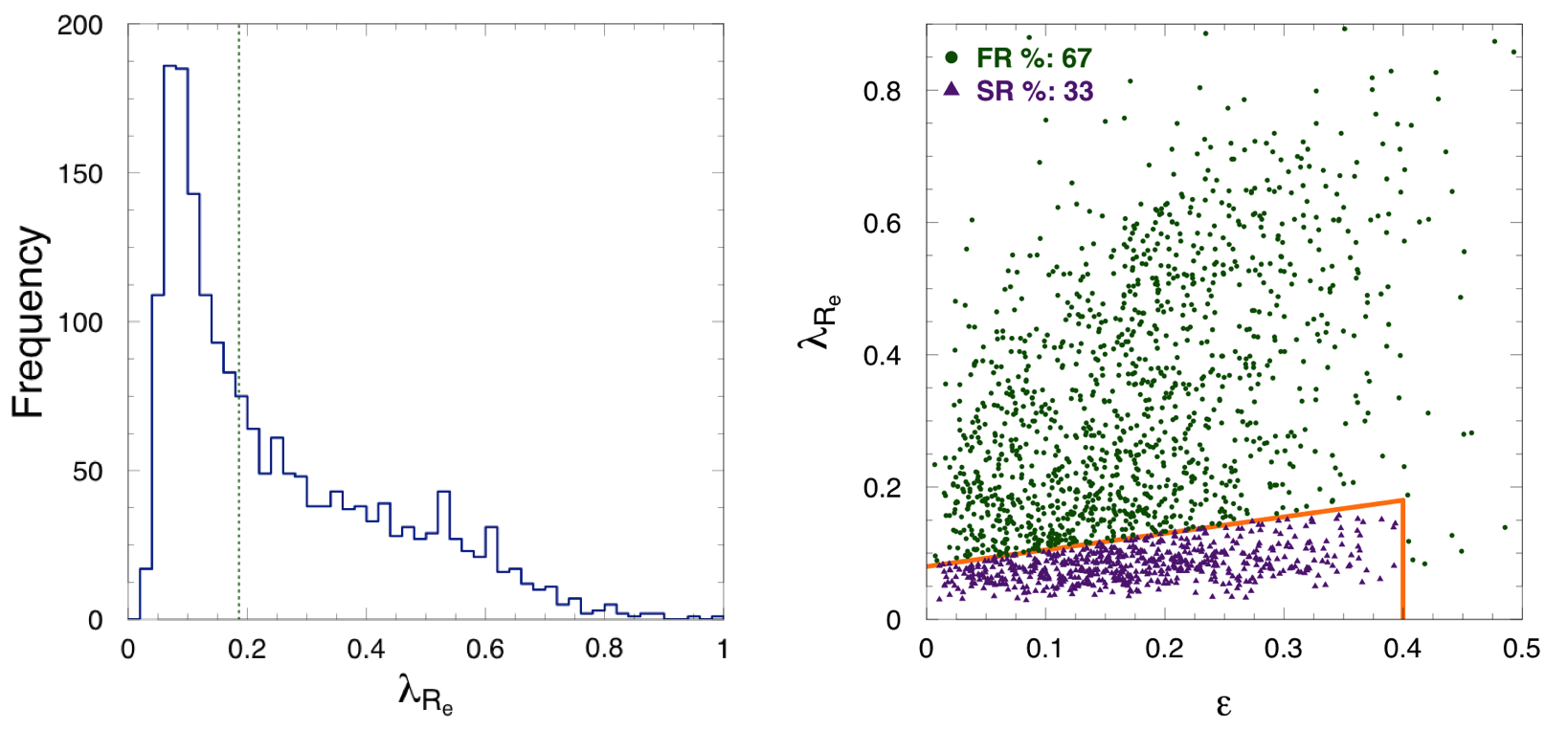} \caption{{\it Left: }Histogram of the  luminosity-weighted stellar angular momentum parameter $\lambda_{\rm Re}$ showing a unimodal distribution with a median of 0.19 (dotted line). {\it Right: }The $\lambda_{\rm Re}$--$\epsilon$ plane and the classification of the stellar kinematics of our  MaNGA sample. The region where SRs are expected to dominate is delineated by the orange lines. EGs located outside of these boundaries are classified as FRs.}
\label{fig:kinematics}
\end{center}
\end{figure*}

The unprecedented quality of  MaNGA's IFU data allows us to determine an important piece of information for characterizing the pattern of the stellar velocity field, the luminosity-weighted stellar angular momentum parameter \citep{Emsellem07}:
\begin{equation}
 \lambda_{\rm Re}\equiv\frac{R|V|}{R\sqrt{V^2+\sigma^2}}=\frac{\sum_{n=1}^NF_nR_n|V_n|}{\sum_{n=1}^NF_nR_n\sqrt{V_n^2+\sigma_n^2}}
\label{eq:lambda.Re} 
\end{equation}
The summation is performed over $N$ spaxels within the radius $R_n$. The quantities $F_n$, $V_n$ and $\sigma_n$ are respectively the flux, projected velocity and velocity dispersion of the $n$th spaxel. We executed the python package \texttt{lambdaR-e-calc} provided by \cite{Graham18}, to calculate $\lambda_{\rm Re}$ within $R_{50}$, given the 2D kinematic data (\texttt{MAPS-HYB10-MILESHC-MASTARSSP}). That resulted in a single-peaked distribution of $\lambda_{\rm Re}$, as indicated by the {\sc dip test} for unimodality \citep[$p$-value~$= 0.99$;][]{dip} and illustrated in Fig.~\ref{fig:kinematics}.

The pattern of stellar rotation also affects the EG's ellipticity \citep[][]{Binney08}, making the combination of  $\lambda_{\rm Re}$ and $\epsilon$ a powerful tool for distinguishing between FRs and SRs. Following \cite{Graham18}, we defined SRs by galaxies lying within the region satisfying $\lambda_{\rm Re} < 0.08 + \epsilon/4$ and $\epsilon < 0.4$ on the $\lambda_{\rm Re}$--$\epsilon$ plane. EGs located outside of this region are identified as FRs (Fig.~\ref{fig:kinematics}).

We found that 68\% of our sample EGs are classified as fast rotators. This fraction is very close to that found by \cite{Emsellem11} (66\% of FRs, although their sample is volume-limited) and also matches the proportion of boxy to disky galaxies, 35\% and 65\%, respectively. Our measurements of $\lambda_{\rm Re}$ also align well with those reported by \cite{Zhu23}, as evidenced by the sharply peaked distribution of the differences between the two estimates at zero ($0.002 \pm 0.067$). We provide the complete catalog of $\lambda_{\rm Re}$ as measured in this work in Appendix~\ref{sec:parameters}.

\section{Exploring the parameter space} 
\label{sec:exploratory}

In the preceding sections, we gathered a diverse set of physical parameters for a large sample of EGs. These measurements not only enable a description of the ongoing activity within the galaxies but also facilitate the search for signatures of physical processes that occurred during the galaxy's lifetime. Our parameter vector,
\begin{equation}
X = \{ \langle a_4/a \rangle, \, {\rm M}_r, \, {\rm M}_\star, \, [{\rm Mg/Fe}], \, \lambda_{\rm Re}, \, \epsilon,  \,  \sigma_\star \} , 
\label{eq:param.vector}
\end{equation}
combines photometric, kinematic, and chemical features of each galaxy, namely the isophotal shape parameter (Section~\ref{sec:light}), the absolute magnitude in the $r$-band and the stellar mass, both measured within the effective radius (Section~\ref{sec:nsa}), the magnesium-to-iron ratio (Section~\ref{sec:chemical}), the luminosity-weighted stellar angular momentum parameter (Section.~\ref{sec:dynamics}), the ellipticity (Section~\ref{sec:nsa}) and the stellar velocity dispersion (Section~\ref{sec:chemical}). Individually, each of these parameters follows a unimodal distribution according to the {\sc dip test} \citep{dip}, with p-values ranging from 0.50 to 0.99.

Having collected such a diverse sample, we are now ready to scrutinize it in search of correlations that may reveal some level of clustering within the data. However, this task is far from straightforward when dealing with a large number of dimensions (seven, in our case) and/or a substantial number of objects (1,895 EGs). To address this complexity and reduce the dimensionality of the data effectively, we employed Principal Component Analysis (PCA), which will be described in detail below.

\subsection{Principal Component Analysis} 
\label{sec:pca}

The Principal Component Analysis (PCA) belongs to the class of statistical techniques known as unsupervised learning. Instead of concerning about making predictions of future observations (as supervised learning does), this class is designed to explore the characteristics of the measured data, by searching for example, for well-defined patterns or sub-groups among the variables.

In short, the goal of PCA is to seek a low-dimensional representation of the data with minimum loss of information. Given a set of features or parameters $X_1,X_2,\dots,X_p$, the $i$-th ($i=1,2,\dots, p$) principal component (PC) corresponds to the normalized linear combination of the parameters $X$:
\begin{equation}
{\rm PC}i=\sum_{j=1}^p \phi_{j, i}X_j,\,  
 \label{eq:PC}
\end{equation}
given the constraint $\sum_{j=1}^p \phi^2_{j,i}=1$. By definition, all of the  resulting $p$ PCs are orthogonal (i.e., independent) to each other. Geometrically, the loading vector $\phi_1=\{ \phi_{11}, \phi_{21}, \dots, \phi_{p1}\}$ of the first principal component (PC1) defines the direction in the original parameter space in which the variance is maximum. Equivalently, the higher the order $i$, the lower the relative variance for each ${\rm PC}_i$. In general, the first two or three components typically can account for most of the sample's total variance, in the sense that one can safely disregard the remaining PCs from the analysis.

Given a data set with $p$ measured parameters labelled $x$ for $n$ observations (or equivalently, an $n\times p$ data matrix {\bf X}), the scores $s_{k,1}$ ($k=1,2,\dots, n$) of the first PC are defined as:
\begin{equation}
s_{k,1}=\phi_{11}x_{k,1}+\phi_{21}x_{k,2}+\dots+\phi_{p1}x_{k,p}\,  
 \label{eq:PC.scores}
\end{equation}
which correspond to the projection of the $n$ data points onto the new axis, referred to as PC1. The scores of other PCs are calculated similarly.

An important caveat is that when there is a significant discrepancy in the variances of the parameters, the PCs may become less informative as each component tends to be dominated by a particular parameter. To mitigate this, we standardized the measurements by subtracting the respective mean and dividing by the standard deviation, $(x_k - \bar{x}_k)/ {\tt std}(x_k)$. We then utilized the {\sc R} function {\sc princomp} to compute the PCs for the MaNGA sample. The summary of the results is presented in Table~\ref{tab:PCs}.

\begin{table*}
\caption{Absolute and relative standard deviation of the principal components (PCs) of our sample of EGs.}
\label{tab:PCs}
\begin{center}
\begin{tabular}{l c c | c c c c c}
\hline
\hline 
& PC1 & PC2 & PC3 & PC4 & PC5 & PC6 & PC7 \\
\hline

Standard deviation     & 1.69 & 1.24 & 0.95 & 0.89 & 0.77 & 0.50 & 0.32 \\
Proportion of Variance & 0.41 & 0.22 & 0.13 & 0.11 & 0.09 & 0.04 & 0.01 \\
Cumulative Proportion  & 0.41 & 0.62 & 0.75 & 0.86 & 0.95 & 0.99 & 1.00 \\

\hline
\hline
\end{tabular}
\end{center}
\end{table*}

The PCA revealed that the majority of the total variance, 62.4\%, is accounted for by PC1 and PC2. Consequently, the dimensionality of our dataset was reduced to just two, where the new variables correspond to the respective scores of the PCs (Eq.~\ref{eq:PC.scores}). Although not directly represented in this new space, the original set of parameters exhibits strong (anti)correlations with either PC1 or PC2, as can be seen in Table~\ref{tab:PCs.corr}. The exception is [Mg/Fe], which shows only a modest correlation with PC1. A graphical representation of PCs at higher orders is presented in Appendix~\ref{sec:pca.higher}.  For completeness, we also present the correlations among each original parameter (Eq.~\ref{eq:param.vector}) in Appendix~\ref{sec:corr}.

\begin{table}
\caption{Correlation of the features to PC1 and PC2.}
\label{tab:PCs.corr}
\begin{center}
\begin{tabular}{l c c}
\hline
\hline 
parameters & corr. w. PC1 & corr. w. PC2 \\
\hline
$\langle a_4/a \rangle$ & -0.29 & -0.56 \\
M$_r$                   & -0.93 & 0.09 \\
M$_\star$               & 0.91 & -0.11 \\
$[{\rm Mg/Fe}]$         & 0.43 & 0.09 \\
$ \lambda_{\rm Re}$     & -0.36 & -0.68 \\
$\epsilon$              & 0.05 & -0.80 \\
$\sigma_\star$          &  0.86 & -0.26 \\
    
\hline
\hline
\end{tabular}
\end{center}
\end{table}

\begin{figure*}
\begin{center}  
\includegraphics[width=1.0\textwidth]{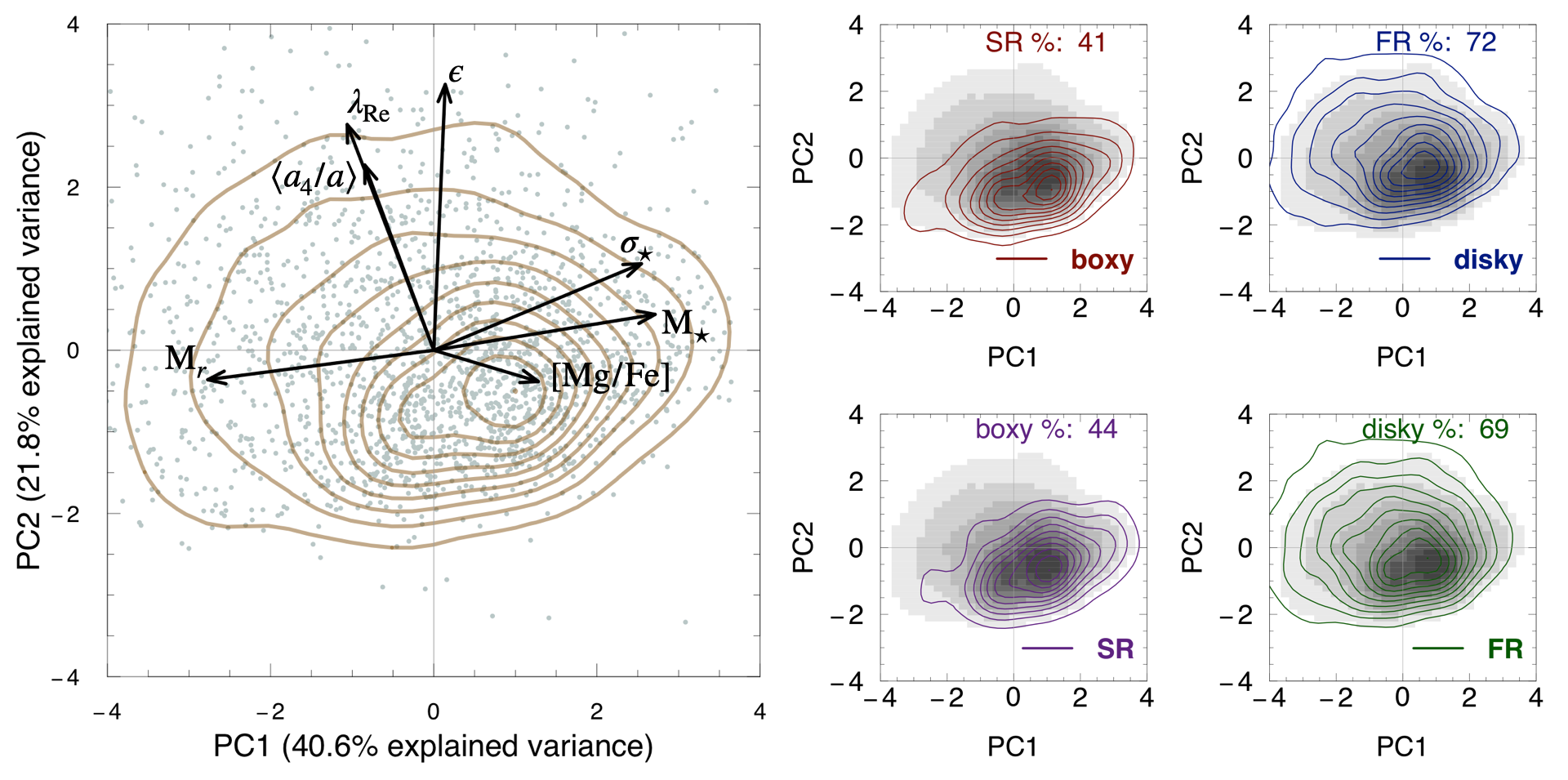} \caption{Scores of the first two principal components, which together account for 62\% of the total sample's variance. {\it Left: } Biplot with loading vectors  representing the contribution of each original parameter to the principal components. Normalized density contours of the smoothed data, starting at 0.1 and rising with increments of 0.1. {\it Right: }Principal component space of four subgroups in our sample. The corresponding scores are represented by colored contours (red contours: boxy EGs; blue: disky EGs; purple: SRs; green: FRs), overlaid with the score distribution of the entire sample (grayscale). The subgroups are based on light distribution (boxy/disky, first row) and stellar kinematics (SR/FR, second row). The percentage of members belonging to another class is shown at the top of each panel (e.g., 41\% of the boxy EGs are also SRs -- see upper left panel).}
\label{fig:pca.biplot}
\end{center}
\end{figure*}

A PC scatter plot is shown on the left-hand side of Fig.~\ref{fig:pca.biplot}. Additionally, we overlaid density contours of the data points smoothed using the function {\sc bkde2d} \citep{KernSmooth}, with a width  of 0.4 in each direction.

The data points form a single clustered pattern, with only a few outliers. In other words, the scores do not segregate into sub-groups in the PC space with distinct characteristics and identifiable boundaries. The implication is that, based on our collection of measurable parameters, the proposed dichotomy scenario of EGs is not supported by the PCA.

In order to investigate the correspondence between the luminosity spatial distribution (disky/boxy) and stellar kinematics (SR/FR) classes, we labeled the corresponding scores in the PC space. The results are depicted in the first two rows on the right-hand side of Fig.~\ref{fig:pca.biplot}. Disky and FRs show a distribution closely matching the general one (gray levels), suggesting these features are representative of the entire sample of EGs. Conversely, boxy and SRs occupy a smaller, yet still overlapping, region, primarily located in the 3rd and 4th quadrants. A consistent shift in the relative composition is observed. The SR/FR ratio in the boxy (disky) subsample is 41/59 (28/72), whereas the overall value is 33/67. The ratio of boxy/disky for SR (FR) is 44/56 (31/69), while for the total sample, it is 35/65.

\begin{figure*}
\begin{center}  
\includegraphics[width=1.0\textwidth]{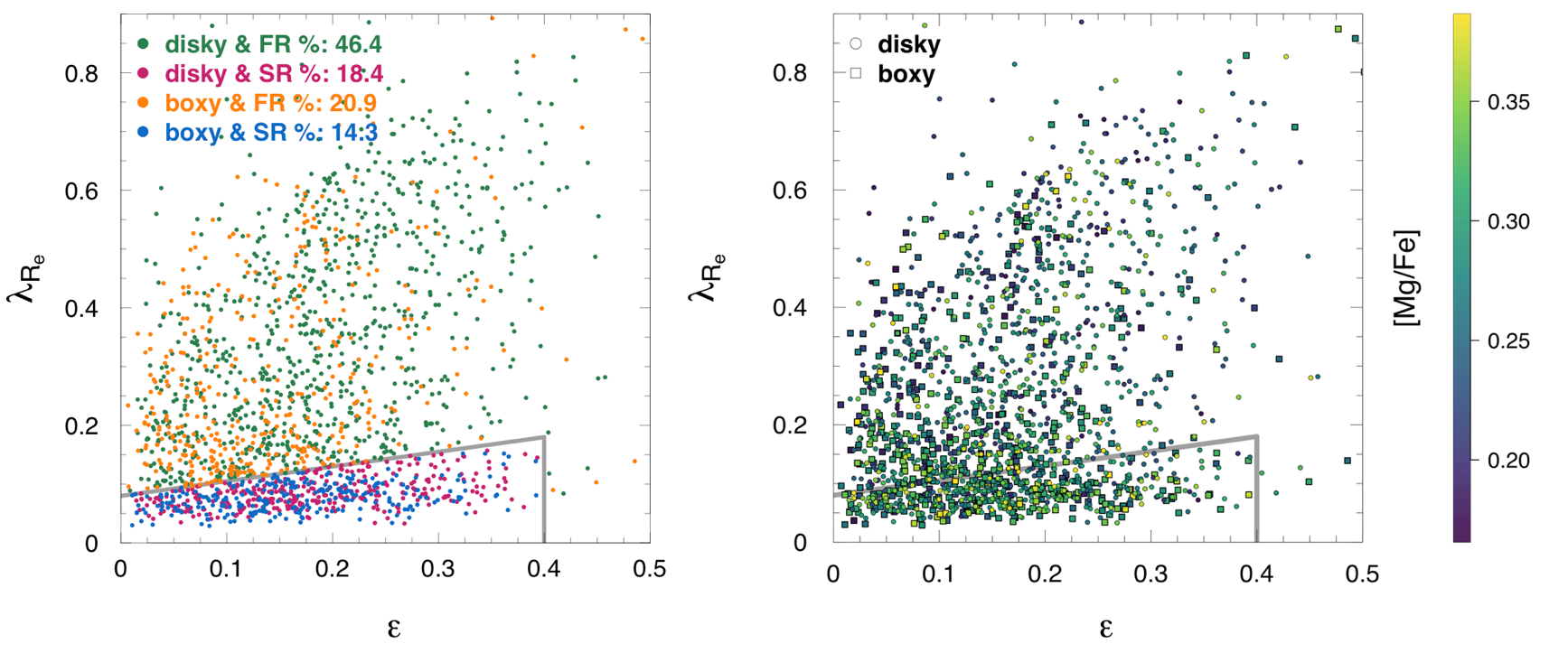} \caption{Distribution of our EGs in the $\lambda_{\rm Re}$--$\epsilon$ plane. {\it Left: } With a two-labeled classification (kinematics \& isophotal shape). {\it Right: } With a three-labeled classification (with the color scale representing the [Mg/Fe] ratio).}
\label{fig:classes}
\end{center}
\end{figure*}

The fractions mentioned above only consider one class at a time per galaxy. The distribution of EGs of different combination of classes is shown in Fig.~\ref{fig:classes}. We confirm that, contrary to the predictions of the dichotomy scenario, there is no one-to-one correspondence between photometric and kinematic classifications. Although similar conclusions have been reached by previous studies \citep[e.g.,][]{Emsellem07,Emsellem11,Cappellari11}, the significantly larger and more homogeneous sample of EGs presented in this work provides unequivocally stronger support for this statement. 
In fact, disky and FRs constitute the most common population, with 46.4\% of the EGs in the MaNGA sample exhibiting these features simultaneously. Following this, boxy and FRs account for 20.9\% of the budget, closely followed by the combination of disky and SRs, which represent 18.4\%. The smallest contribution, 14.3\%, is attributed to boxy and SRs. Notably, this ranking remains unchanged (despite a subtle shift in the relative proportions) when considering different subsets of stellar velocity dispersion, as shown in Table~\ref{tab:fractions}. No clear trend with the [Mg/Fe] ratio is observed, as shown in the right-hand panel of Fig.~\ref{fig:classes}.

\begin{table*}
\caption{Different combinations of classes as a function of stellar velocity dispersion.}
\label{tab:fractions}
\begin{center}
\begin{tabular}{l | c c c c c }
\hline
\hline 
& $50\leq \sigma_\star< 100$ & $100\leq \sigma_\star< 150$ & $150\leq \sigma_\star< 200$ & $200\leq \sigma_\star< 250$ & $250\leq \sigma_\star< 500$ \\
& (km \ s$^{-1}$) & (km \ s$^{-1}$) & (km \ s$^{-1}$) & (km \ s$^{-1}$) & (km \ s$^{-1}$) \\
\hline
disky \& FR & 58.8  &  62.7 & 41.2 & 44.5 & 45.1\\
disky \& SR & 13.2  &  8.8  & 20.9 & 19.9 & 19.0\\
boxy \& FR  & 19.3  &  20.2 & 20.4 & 20.7 & 23.3\\
boxy \& SR  &  8.8  &  8.3  & 17.5 & 14.9 & 12.6\\
\hline
Number of EGs & 114 &  228 & 542 & 685 & 326\\
\hline
\hline
\end{tabular}
\end{center}
\end{table*}

Our exploratory analysis has indicated that the transition among the properties of EGs is smoother than presumed by the dichotomy. However, this does not preclude the possibility that the population of EGs may exhibit some trends, particularly when considering the extreme values of specific parameters. Consequently, a noticeable change in the relative composition might occur, potentially leading to a false dichotomy. In the following sections, we will delve into quantifying these trends involving the photometric and kinematic classes, as well as the remaining parameters.

\subsection{Boxy vs.~Disky Fraction} 
\label{sec:disky.boxy}

The PCA has revealed that our sample of EGs forms an undeniably single cluster within the PC space. In other words, the boxy/disky or SR/FR classes lack clearly defined boundaries that would enable unambiguous classification of a galaxy if one of these features is absent.

However, as illustrated in Fig.~\ref{fig:pca.biplot}, different classes are not evenly distributed in the PC-space. While the fourth quadrant harbors the largest number of galaxies, a significant proportion of disky EGs or FRs can be found outside of this region, particularly in the first and second quadrants, where the density of boxy EGs, as well as SRs, is comparatively lower. This asymmetric distribution suggests a potential variation in the proportion of boxy/disky or SR/FR concerning other parameters -- $M_r$, $\sigma_{\star}$, M$_\star$, and [Mg/Fe].

To ensure the robustness of our results and minimize any bias due to potential misclassification, we constructed a subsample comprising galaxies identified as EGs by both MaNGA and GZ2 (Fig.~\ref{fig:sample}). Additionally, we only included galaxies classified within a 99\% c.l.~threshold. This stringent criterion ensures that even considering the most extreme possible values of parameters, such as $\overline{\langle a_4/a \rangle}\pm3\sigma_{\langle a_4/a \rangle}$, where $\sigma_{\langle a_4/a \rangle}$ represents the standard deviation, a galaxy remains within the same class. Following these stringent criteria, our subsample consisted of 621 galaxies.

Now, we focus on examining how the fraction of boxy or disky EGs, depends on key galaxy properties, specifically ${\rm M}_r$, $\sigma_\star$, and ${\rm M}_\star$. Each fraction was determined relative to the total number of galaxies within a specific bin, ensuring that the sum equaled to one. Error bars were computed from the underlying binomial distribution using the {\sc Rbeta.inv} function \citep{zipfR}. We assessed three distinct models to elucidate the correlation between the fraction of galaxies and a given feature (${\rm M}_r$, $\sigma_\star$, or ${\rm M}_\star$):
\begin{itemize}
    \item  Model~\#1 (monotonic): a single linear fit with $\theta = \{a, b\}$ (see below for the definition of $\theta$);
    \item  Model~\#2 (monotonic with non-constant slope): a double linear fit with a inflection point, $\theta = \{a_1, a_2, b_1, b_2 \}$; and
    \item  Model~\#3 (uniform distribution): a constant fit equals to the weighted mean, $\theta = \{b_3\}$.
\end{itemize}
In each case, $\theta$ represents the set of the model's free parameters, where $a$ and $b$ represent the slope and intercept of the linear function(s), respectively. 
The $\chi^2$ statistic for each model is expressed as \citep{Wall12}:
\begin{equation}
\chi^2=\sum_{i=1}^{n}  \frac{(F_i-f_i)^2}{\sigma_{\rm low}^2 + \sigma_{\rm upp}^2}  . 
\label{eq:chi2.shear}
\end{equation}
Here, $F_i$ represents the fraction predicted by the model, $f_i$ denotes the measured fraction, and $\sigma_{\rm low}$ and $\sigma_{\rm upp}$ are the lower and upper error bars for each of the $n$ observations. The corresponding likelihood is $\mathcal{L} \propto \exp(-\chi^2/2)$. We then mapped each posterior ${\rm Pr}(\theta|{\rm data}) \propto \mathcal{L}({\rm data}|\theta)$ using the Markov Chain Monte Carlo method, sampled by the random walk algorithm Metropolis \citep[{\sc MCMCmetrop1R};][]{MCMCpack}. For each model, a chain of $10^4$ elements preceded by $10^3$ iterations for burning-in was generated.

The choice of the best model was determined using the Bayesian Information Criterion,
\begin{equation}
    {\rm BIC}=k\ln{n}-2\ln{\mathcal{L}_\star} \,.
    \label{eq:bic}
\end{equation}
This index simultaneously considers the number  of parameters in the model ($k$), the number  of data points ($n$), and the maximum log-likelihood $\mathcal{L}_\star$. The BIC is preferred over other indices \citep[such as the Akaike Information Criterion; e.g.,][]{Monteiro-Oliveira22b} since it penalizes more complex models, i.e., those with a large number of parameters, while also taking into account the number of data points. Among a set of finite models, the one with the lowest BIC is considered the best description of the data. However, model assessment may not be straightforward, especially when the BIC statistics are similar. As a guideline, \cite{kass95} offer a rule of thumb for comparing two models, A and B, which is presented in Table~\ref{tab:bic}.

\begin{table}
\caption{Guidelines for model comparison based on the BIC statistics are presented in Table~\ref{tab:bic}. In this example, model A has the lowest BIC value}
\label{tab:bic}
\begin{center}
\begin{tabular}{c c}
\hline
\hline 
$\Delta{\rm BIC}$ & comment \\ 
$({\rm BIC}_{\rm B} - {\rm BIC}_{\rm A})$  \\
\hline

\hline
$\leq 2$ &  A and B comparable\\
2 -- 6 & a positive evidence in favor of  A\\
6 -- 10 & strong evidence in favor of  A\\
$> 10$ & very strong evidence in favor of  A\\
\hline
\end{tabular}
\end{center}
\end{table}

\begin{figure}
\begin{center} 
\includegraphics[width=\columnwidth]{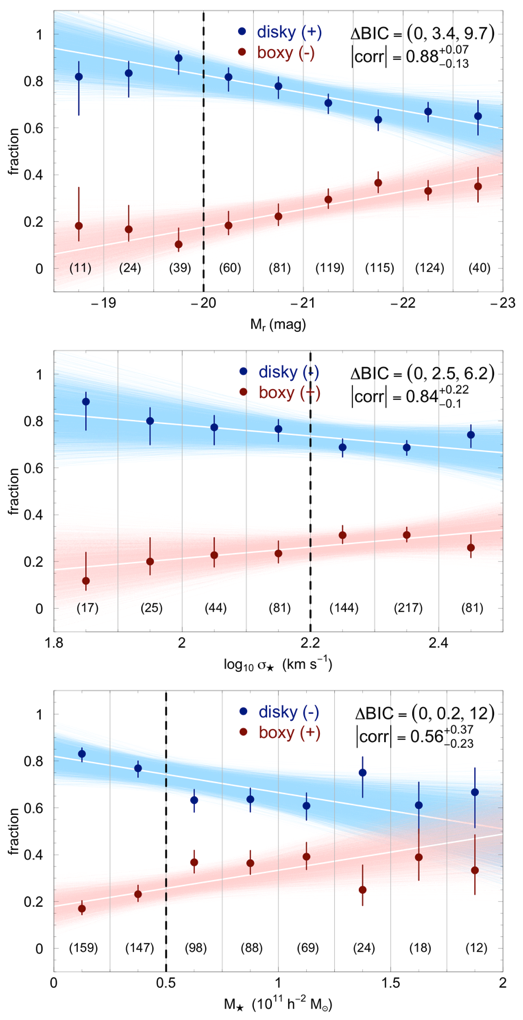} \caption{Fraction of disky and boxy galaxies as a function of absolute magnitude (upper panel), stellar velocity dispersion (middle panel), and stellar mass (lower panel). The total fraction within each bin containing $N$ galaxies is equal to one. In these plots, we considered a subset of 621 EGs matching the MaNGA and GZ2 catalogs, ensuring that the classification as boxy or disky remains unchanged if any value of $\langle a_4/a \rangle$ falls within $\overline{\langle a_4/a \rangle}\pm3\sigma_{\langle a_4/a \rangle}$. The sign next to the labels indicates whether the respective Pearson's correlation coefficient shown in the right-hand corner of each panel is positive or negative. The legend also presents the model assessment based on the $\Delta$BIC statistics computed from models \#1 (linear fit with constant slope), \#2 (linear fit with non-constant slope), and \#3 (constant fraction), respectively. The white line represents the best single linear model, with its 68\% c.l. shown by the colored shadow. The vertical dashed line indicates the chosen inflection point (for model 2).}
\label{fig:fraction.bd}
\end{center}
\end{figure}

The variation of the boxy/disky fraction is shown in Fig.~\ref{fig:fraction.bd}. The bin size is chosen to ensure a minimum of at least ten galaxies. Two pieces of evidence indicate that the fractions strongly depend on the absolute magnitude (top panel), stellar velocity dispersion (middle panel), and stellar mass (lower panel): the high absolute value of the Pearson's correlation coefficient and the rejection of the constant fraction model. Despite models \#1 and \#2 being statistically equivalent, the principle of Occam's razor strongly favors the simpler one (\#1). Notably, the results remain consistent even with an increase in the bin width or a change in the choice of the inflection point (for model \#2).

The single linear fit aligns with the prescription of the dichotomy scenario, where the disky fraction decreases as galaxy luminosity increases, while the opposite behavior holds for boxy galaxies. The most pronounced contrast is observed in the faintest branch, dominated by disky galaxies, which represent approximately 85\% of the total counts. Conversely, both fractions converge to around 50\% in the brightest branch. Similar trends are observed when considering stellar velocity dispersion and stellar mass. Importantly, these conclusions hold true when considering our entire MaNGA sample.

\subsection{SR vs.~FR Fraction} 
\label{sec:FR.SR}

\begin{figure}
\begin{center} 
\includegraphics[width=\columnwidth]{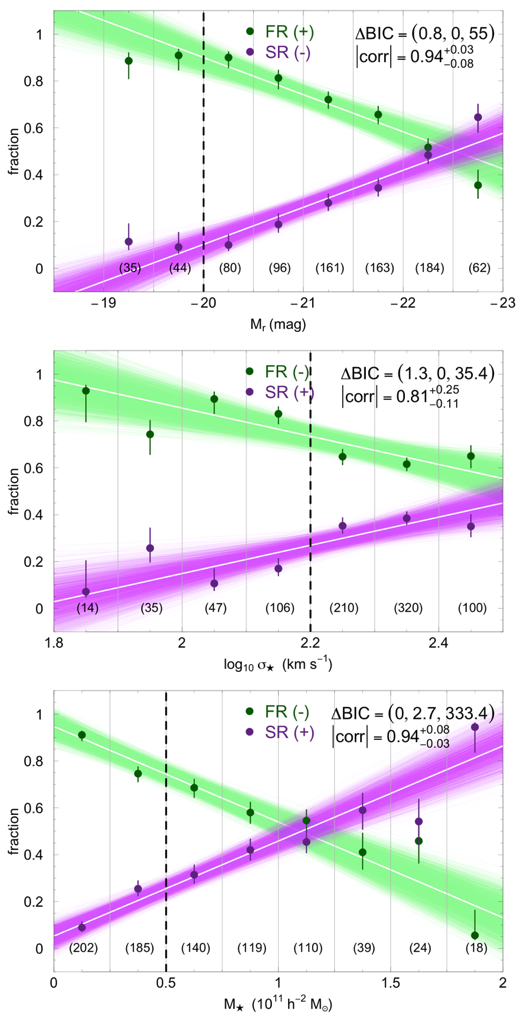} \caption{Similar to Fig.~\ref{fig:fraction.bd}, but for the fraction of FR and SRs of those EGs in common between MaNGA and GZ2 and matching a kinematic classification with 99\% c.l.}
\label{fig:fraction.sf}
\end{center}
\end{figure}

We defined a subset of 845 FRs and SRs in a similar manner as in the previous section, encompassing those classified within the 99\% c.l. to ensure they are safely distant from the boundaries defined in Fig.~\ref{fig:kinematics}. The SR/FR fractions are depicted in Fig.~\ref{fig:fraction.sf} along with the respective best linear fits.

Overall, one can observe the same pattern as for the fractions of disky and boxy galaxies, although there are noticeable differences. None of the 21 galaxies common between MaNGA and GZ2 and fainter than $M_r = -19$ were selected after applying our restrictive criteria. For the sake of comparison, 70\% of the 89 galaxies originally within $-23 \leq M_r \leq -22.5$ were selected. This suggests that reliable measurements of kinematics for fainter galaxies are still noisy even in the high-quality data of the MaNGA survey.

Regarding the relative composition, unlike boxy EGs, SRs become the dominant type in the rightmost bins, both in luminosity and stellar mass. This corresponds to a limit of $M_r\lesssim -22.5$ and $M_\star\gtrsim 1.25\times 10^{11} h^{-2} M_\odot$, respectively. It is within this range that one finds the brightest cluster galaxies \citep[BCGs; e.g.,][]{Dalal21,Zenteno25}, usually located at the bottom of the potential well of galaxy clusters \citep[e.g.,][]{Machado15b,Soja18,Pandge19,Monteiro-Oliveira17a,Monteiro-Oliveira17b,Monteiro-Oliveira18,Monteiro-Oliveira20,Doubrawa20,Kelkar20,Moura21,Monteiro-Oliveira21,Monteiro-Oliveira22a,Hernandez-Lang22,Monteiro-Oliveira22b,Albuquerque24,Machado24,Ding25}. As suggested in Fig.~\ref{fig:fraction.sf}, these galaxies exhibit distinct properties compared to the bulk of EGs, possibly due to their co-evolution with the host cluster \citep[e.g.,][]{Lin04}, although further details remain unclear.

\subsection{Magnesium-to-Iron Ratio} 
\label{sec:chemical.fraction}

\begin{figure*}[ht!]
\begin{center}
\includegraphics[width=\textwidth]{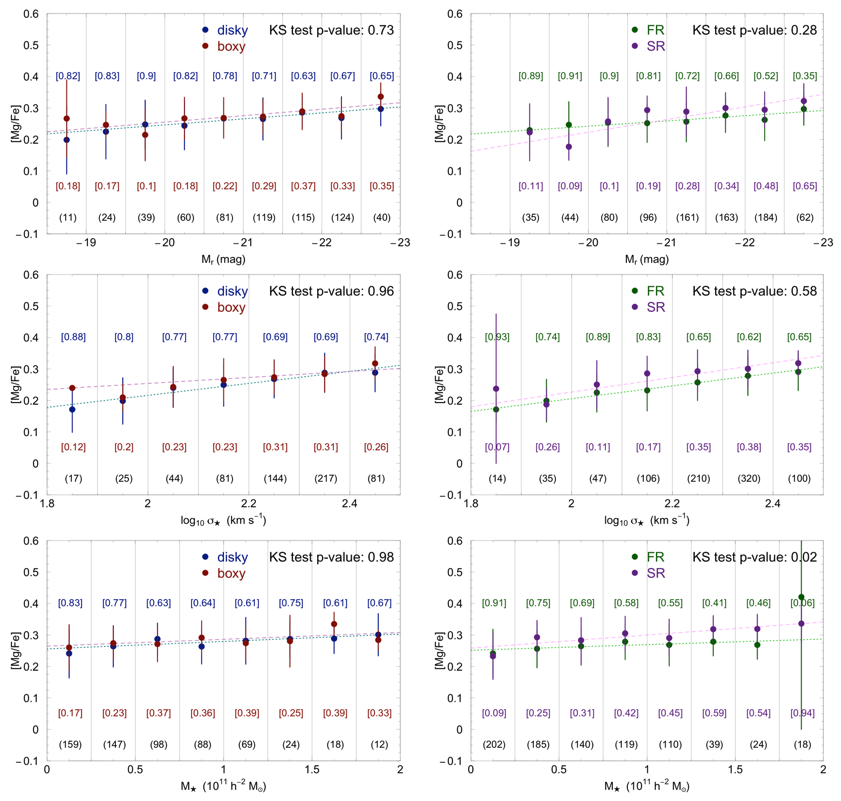} \caption{Magnesium-to-iron ratio for the photometric and kinematic classes. Each sample was selected following the criteria described in Sec.\ref{sec:disky.boxy} and Sec.\ref{sec:FR.SR}, respectively. The numbers in brackets show the respective contribution for each bin, following the color code in the legend. The $p$-value refers to the result of the Kolmogorov–Smirnov (KS) test.}
\label{fig:mgfe}
\end{center}
\end{figure*}

We end the parameter space exploration by investigating the behavior of the [Mg/Fe] ratio for each of the EGs classes. Here, we use the same subsets as defined in Sections~\ref{sec:disky.boxy} and~\ref{sec:FR.SR}, respectively, for the photometric and kinematic classifications. The results are presented in Fig.~\ref{fig:mgfe} with the same bin sizes used before.

Regardless of the labels/classifications, the mean values of each bin (taken as the corresponding mean of [Mg/Fe]) are very close to each other. This, coupled with the high variance, leads us to conclude that the $\alpha$-to-iron ratio does not depend on the different isophotal or kinematic classifications. This conclusion is supported by the Kolmogorov-Smirnov tests \citep[{\sc ks.test};][]{R}, which, for all cases except the comparison between kinematics and stellar mass, show that the two corresponding samples are statistically equivalent ($p$-value~$\geq 0.05$).

\section{Discussion} 
\label{sec:discussion}

\subsection{Robustness of the Analysis}

A potential concern for our analysis is the contamination by S0 galaxies in our EG sample. Given their inherently substantial angular momentum, it is not expected that S0s are abundantly found in the region occupied by SRs. However, their presence can coincide with that of FRs in the  $ \lambda_{\rm Re}$--$\epsilon$ plane. As illustrated by \cite{Graham18} in their Fig.~8, the majority of EGs are entirely contained within $\epsilon \leq 0.4$ and $ \lambda_{\rm Re} \leq 0.8$, with reasonably low contamination by other morphological types. In contrast, S0s tend to exhibit greater elongation ($\epsilon>0.4$) and possess higher angular momentum ($\lambda_{\rm Re}\geq 0.8 $). In our sample, only 39 galaxies (2\% of the total MaNGA sample) satisfy both criteria simultaneously, suggesting that potential contamination by S0s is negligible. Consequently, the identification of EGs by the machine learning method implemented in MaNGA can be considered trustworthy. Another piece of evidence supporting our confidence in the high purity of the EGs sample is that the proportion of FRs found in this work (67\%, Fig.~\ref{fig:kinematics}) is equivalent to those found by \cite{Emsellem11} (66\%) among the EG sample in the ATLAS$^{\rm 3D}$ project.

The Sérsic index ($N_s$), derived from a two-dimensional, single-component Sérsic fit in the r-band available in the NSA, provides additional evidence supporting the purity of our sample. In fact, 98\% of our sample has $N_s>3$, with a median value of $N_s=6$. This sharp distribution does not offer any improvement when included in the parameter vector. Notably, there is no significant change in the first two principal components, except for a slight decrease in their contribution to the total variance, as expected due to the addition of a new dimension.

The stellar kinematics characterization can be dramatically affected by the galaxy's viewing angle $\Theta$ (or inclination). The closer it is to being perpendicular to the plane of the sky (i.e., $\Theta\approx 0$ for a face-on view), the greater the uncertainties in the kinematic parameters, which could potentially lead to erroneous classification and introduce bias in the PCA. To verify possible systematic errors introduced in the PCA by the inclusion of low-$\Theta$ galaxies, we utilized the catalog of \cite{Zhu23}, containing dynamic and kinematic properties of the entire MaNGA survey sample obtained from the Jeans Anisotropic Modeling (JAM) method.  Among the collection of models provided by the authors, we adopted the one where both the dark and luminous matter distributions are equivalent. Then, all galaxies with $\Theta<30$ degrees were excluded from our sample. The remaining 1585 galaxies exhibit exactly the same SR/FR ratio as our original MaNGA sample (Fig.~\ref{fig:kinematics}), and subsequent PCA yielded equivalent results to those described in Section \ref{sec:pca} and illustrated in Fig.~\ref{fig:pca.biplot}. 
Additionally, adopting more restrictive cuts ($\Theta < \{60, 80\}$ degrees), or employing different dynamical models does not lead to significant changes in our PCA results.
These findings align with the conclusions of \cite{Emsellem11} and \cite{Graham18}, who argued that the kinematic classification based on the $\lambda_{\rm Re}$--$\epsilon$ plane has low dependence on the viewing angle. Our conclusion regarding the unimodal distribution of EGs in the PC space remains robust even when adopting kinematic parameters measured by \cite{Zhu23} (the luminosity-weighted stellar angular momentum parameter, and the ellipticity, respectively, {\sc Lambda\_Re} and {\sc Eps\_MGE}) for kinematic classification. Regarding the photometric classification, it is worth mentioning that while the isophotal shape might be influenced by projection effects, the underlying signal of $a_4/a$ (Eq.~\ref{eq.a4}) remains unaffected \citep{Bender89}.

\subsection{Absence of a Dichotomy?}

Our comprehensive analysis unambiguously indicates the absence of a dichotomy among EGs, at least among the properties we considered (namely the isophotal shape, absolute magnitude, stellar mass, [Mg/Fe], stellar kinematics, and  stellar velocity dispersion). In other words, knowing a galaxy to be slowly rotating and having boxy isophote does not imply a high [Mg/Fe] value. Similarly, an EG that is fast rotating and having a low [Mg/Fe] ratio can still have a boxy isophote.

Overall, the majority of EGs exhibit disky-shaped isophotes and stellar kinematics consistent with a FR pattern. This characterization holds simultaneously for approximately 46\% of the MaNGA sample, but it can increase to 65\% for EGs fainter than $M_r=-20$ or even 69\% at $M_\star\leq 2.5\times 10^{10} M_\odot$. Both diskys and FRs are distributed across all quadrants of the PC space (see Fig.~\ref{fig:pca.biplot}), indicating that their properties span a wide range of values. Conversely, boxys and SRs, while appearing in smaller proportions, seem to constitute distinct classes, sparsely populating the upper part of the PC space. Although the fraction of boxy EGs never exceeds that of disky ones, SRs are found to surpass the fraction of FRs both at the brightest branch ($M_r\leq-22.5$, 53\% versus 47\%) and at the high-mass end ($M_\star\geq 1.25\times 10^{11} M_\odot$, 64\% versus 36\%), when we consider the 99\% c.l.~sample (Section~\ref{sec:FR.SR}).

Our findings unequivocally demonstrate that neither the disky/boxy nor FR/SR classes exhibit discernible differences in their $\alpha$-to-iron ratio, as inferred from the [Mg/Fe] abundance measured within $R_{50}$. Consequently, we confidently refute the hypothesis proposed by \cite{Kormendy09} that [Mg/Fe] segregates the EGs into two distinct groups. It is noteworthy, however, that their assertions find limited support within a small fraction of EGs, comprising only 2\% of our sample, as depicted in their Fig.~45. In a more recent investigation, \cite{Bernardi23} similarly suggested, based on stacked spectra from MaNGA, that [Mg/Fe] might exhibit subtle dissimilarities between SR and FRs. However, their conclusion was drawn from a narrow range of magnitudes ($-23.5\leq M_r\leq -21.5$), masking the broader trend that shows a gradual transition of [Mg/Fe] across a wider magnitude range, as illustrated in the upper right panel of Fig.~\ref{fig:mgfe}.

The outcome of our work also connects to the long-standing discussion of whether bright (giant) and faint (dwarf) EGs constitute two distinct families or instead form a continuous sequence--a debate that gained renewed attention primarily from the 1980s through the early 2000s. Classical studies such as \cite{Kormendy85, Kormendy89, Binggeli91, Ferguson94}, reviewed evidence in favor of a dichotomy based on structural and photometric differences. However, later works, including \cite{Graham03, Gavazzi05, Cote06}, challenged this view by demonstrating a more continuous distribution of galaxy properties across a wide luminosity range. Although our sample primarily includes brighter systems, our findings align with this broader perspective, reinforcing the view that EGs form a continuum in both their photometric and kinematic properties, rather than falling into clearly distinct classes.

\subsection{Radio Activity}

Our previous analyses have established that the selected EGs' properties form a continuous sequence. However, does the same conclusion extend to wavelengths other than optical? \cite{Bender89} and \cite{Kormendy09} argued that the dichotomy can also be observable in X-ray and radio activities, with a more pronounced effect in boxy and SRs. According to \cite{Kormendy09} and references therein, the mechanism behind the increase in emission at both ends of the electromagnetic spectrum is related to the gas heating process through radio AGN feedback. Supported by a robust theoretical and computational foundation, the authors also proposed that there must be a clear transition between the two regimes (radio/X-ray loud and quiet) around $M_{\star, \rm lim}\simeq 10^{11} M_\odot$. This characteristic mass scale is a direct outcome of the galaxy formation process. In systems exceeding this limit, the strong gravitational potential of the galaxy implies the accretion of gas from the large-scale structure governed by more chaotic processes. This generates shock waves that heat the gas, consequently quenching star formation. The heating is then sustained by the aforementioned AGN mechanism.

Aiming to ascertain if the dichotomy may show up beyond the optical domain, we augmented our parameter vector (Eq.~\ref{eq:param.vector}) with an additional piece of information: the 1.4 GHz radio luminosity ($P_{1.4}$, in W\,Hz$^{-1}$) obtained from the volume-limited catalog of radio AGNs compiled by \cite{Lin18}. This catalog comprises a homogeneous compilation of optical sources from either the SDSS DR6 \citep{Lin10} or DR7 \citep{Best12}, and their counterparts in the 1.4 GHz radio source catalog from the NRAO VLA Sky Survey (NVSS) or Faint Images of the Radio Sky at Twenty-centimeters (FIRST). Furthermore, the sources in \cite{Lin18} were restricted based on specific criteria: $z\leq0.15$, $M_r^{0.1}\leq-21.57$ (i.e., the SDSS $r$-band shifted blueward by a factor of 1.1 in wavelength), and $23.3\leq \log( P_{1.4})\leq 25.5$. This selection process resulted in a total of 2261 radio galaxies, of which 250 (13\%) were in common with our full sample. The PCA applied to this radio-loud subsample supports our previous findings: the resulting PCA scores (Fig.~\ref{fig:match.ytl.pca}) exhibit a unimodal distribution, thereby refuting the hypothesis that radio activity splits the set of EGs into two distinct groups.

\begin{figure}
\begin{center}
\includegraphics[width=1.0\columnwidth]{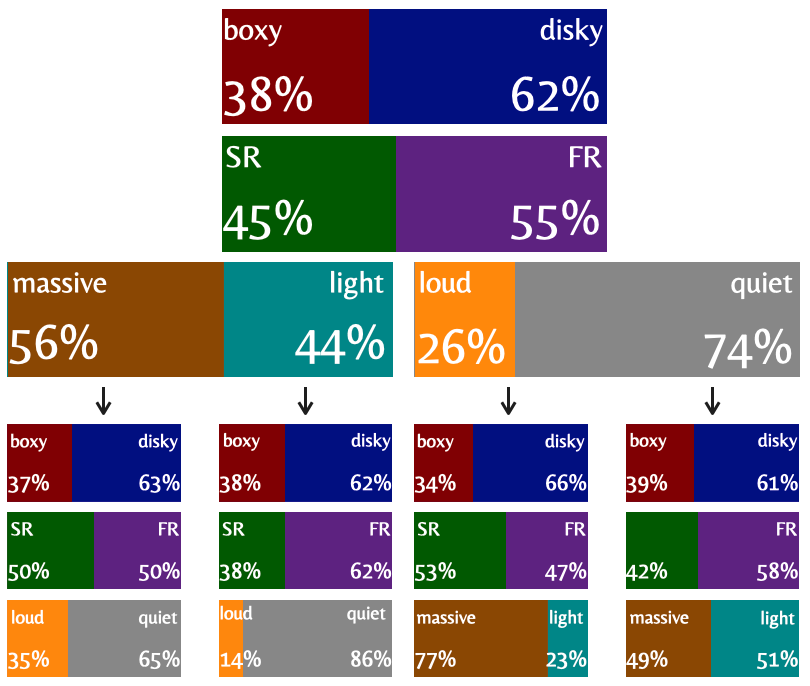} \caption{Statistical summary of the radio sub-sample, initially consisting of 250 EGs. In addition to the classification based on isophotal shape and kinematics, we introduced new categories based on stellar mass (massive/light) and radio activity (loud/quiet).  Please note that even for the ``quiet'' EGs, their 1.4 GHz radio power $P_{1.4}$ (in unit of W/Hz) is still in the range $23.3 \le \log P_{1.4} \le 24$ and thus are still most likely due to super massive black holes, rather than star formation activity.}
\label{fig:match.ytl}
\end{center}
\end{figure}

\begin{figure}
\begin{center}
\includegraphics[width=\columnwidth]{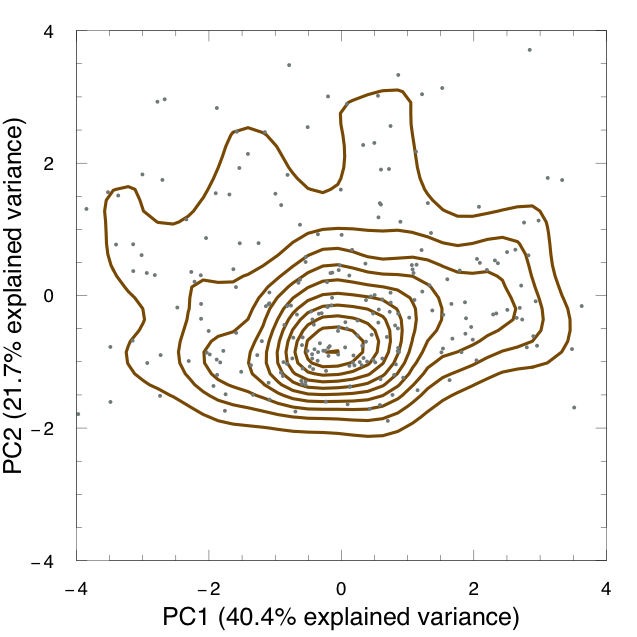} \caption{Scores of the first two PCs of the radio subsample (250 galaxies). The parameter vector encompasses 8 elements: the same listed in Eqn.~\ref{eq:param.vector} plus the 1.4 GHz radio luminosity.}
\label{fig:match.ytl.pca}
\end{center}
\end{figure}

Although no groundbreaking behavior was found in the radio subsample, it is still worth checking for trends among different classes (disky/boxy and FR/SR), as previously performed. Here, we introduced two new labels based on stellar mass: ``massive'' if $M_\star\geq M_{\star, \rm lim}$ or ``light'' otherwise, and radio activity: ``loud'' if $\log P_{1.4} \geq 24$ or ``quiet'' otherwise. The statistics, presented in Fig.~\ref{fig:match.ytl}, reveal a higher proportion of SRs (45\% versus 33\%) and the massive ones (56\% versus 30\%) compared to the MaNGA sample. On the other hand, the fraction of boxy/disky remains roughly the same. Notably, only 26\% of the subsample comprises radio-loud EGs. However, as shown in Fig.~\ref{fig:match.ytl}, it becomes evident that radio activity, when present, prevails in the more massive EGs, which is consistent with numerous previous investigations \citep[e.g.,][]{Best05,Lin07}. Despite this trend, the PCA points out that the transition to the radio-quiet regime is remarkably smooth (Fig. \ref{fig:match.ytl.pca}).

Our findings challenge the notion of a overly simplistic model for the formation of EGs \citep[e.g.,][]{Naab99}. The existence of a myriad of combinations among EG properties, particularly those related to light distribution (isophotal shape) and stellar kinematic patterns, suggests that galaxy assembly results from diverse pathways. In this scenario, distinct matter accretion processes (i.e., major/minor mergers) occurred at different epochs, each involving varying levels of gas contribution. The existence of multiple merger trees is supported by numerical simulations of galaxy formation within the cosmological context \citep[e.g.,][]{Naab14, Penoyre17}, from which the diversity of EGs emerges naturally. However, there is still a lack of better understanding regarding whether certain present-day features represent a transient phase rather than a long-term state. For example, one in every five EGs in our sample exhibits disk-like kinematics simultaneously with boxy-shaped isophotes. Since ordered orbit stars are expected to elongate the light distribution towards the main axis, this combination appears to confront our theoretical framework for describing galaxy kinematics and its connection with light distribution. Understanding this question, which is beyond the scope of this paper, would benefit from follow-up numerical simulations on the evolution of such a group of EGs.

In this study, a key parameter notably absent for assessing the dichotomy scenario, as highlighted by \cite{Kormendy09}, is the central light distribution, which is not available for our sample. Future data from space-based telescopes, such as Roman, should be collected to further investigate the lack of support for the dichotomy scenario identified in this work.

\section{Summary} 
\label{sec:summary}

We have constructed a comprehensive catalog containing seven different physical properties for a large sample of 1895 elliptical galaxies. This catalog includes parameters compiled from other sources, such as the absolute magnitude in the $r$-band and stellar mass, as well as new measurements such as the isophotal shape parameter (allowing classification into boxy or disky classes), magnesium-to-iron abundance ratio, stellar velocity dispersion, luminosity-weighted stellar angular momentum parameter, and galaxy ellipticity. The last two parameters are fundamental for characterizing the stellar rotation pattern as slow or fast. The main findings of our work are listed below:

\begin{itemize}

    \item The majority of EGs in our sample (46.4\%) are FRs and exhibit disky isophotes.

    \item Our principal component analysis indicates that EGs cannot be divided into two distinct groups, and therefore there is no dichotomy. 

    \item As a consequence, isophotal shapes (or any other feature) cannot be used as a proxy for EG kinematics.

    \item Although no distinguishable groups exist, EGs exhibit certain trends with luminosity, stellar velocity dispersion, and stellar mass.

    \item The existence of diverse combinations of properties among the EGs aligns with a complex formation scenario, which encompasses various evolutionary pathways.

\end{itemize}

\begin{acknowledgments}
We acknowledge support from the National Science and Technology Council of Taiwan under grants MOST 111-2112-M-001-043, NSTC 111-2628-M-002-005-MY4,  NSTC 112-2112-M-001-061, and NSTC 113-2112-M-001-005.
RMO is grateful to YTL for all the inspiration, mentoring, and encouragement. This gratitude is also extended to the ASIAA staff and members for providing a pleasant and enriching work environment.
RMO sincerely thanks GMS for all the support, inspiration, and countless Little things that will always be cherished, no matter where we are (6222, 24.13435, 120.67942), Kadu Barbosa for his valuable advices on spectra modelling,  Ting-Wen Lan for sharing the code to download the data from SDSS DR14,
Anna Gallazzi for sharing her catalog of [$\alpha$/Mg] measurements, Connor Bottrell for sharing mock images from TNG, and Meng Gu and Charlie Conroy for their critical help with {\tt alf}.
YTL thanks IH, LYL and ALL for constant encouragement and inspiration.

Funding for the Sloan Digital Sky Survey IV has been provided by the Alfred P.~Sloan Foundation, the U.S.~Department of Energy Office of Science, and the Participating Institutions. SDSS acknowledges support and resources from the Center for High-Performance Computing at the University of Utah. The SDSS web site is www.sdss4.org.

\end{acknowledgments}

\appendix

\section{{\tt alf} performance tests}
\label{sec:alf.bench}

For a random sub-sample of 20 EGs, we obtained the final posterior for all parameters in both simple and full modes using two different setups (i.e., $N_{\rm w}$, $N_{\rm b}$, $N_{\rm s}$). The ``cheap'' configuration was set to the default values (256, 2000, 100), while for the ``expensive'' setup, we followed \citet{Conroy18} and used (512, 20000, 1000). We observed that, for a given set of $N_{\rm w}$, $N_{\rm b}$, and $N_{\rm s}$, the execution time was similar regardless of the mode, approximately 6 minutes for the cheap setup and 85 minutes for the expensive setup, based on parallel execution across 18 cores of a high-performance computer (Intel(R) Xeon(R) Silver 4114 CPU @ 2.20GHz, 125 Gb RAM).

\begin{figure}
\begin{center}
\includegraphics[width=1.0\textwidth]{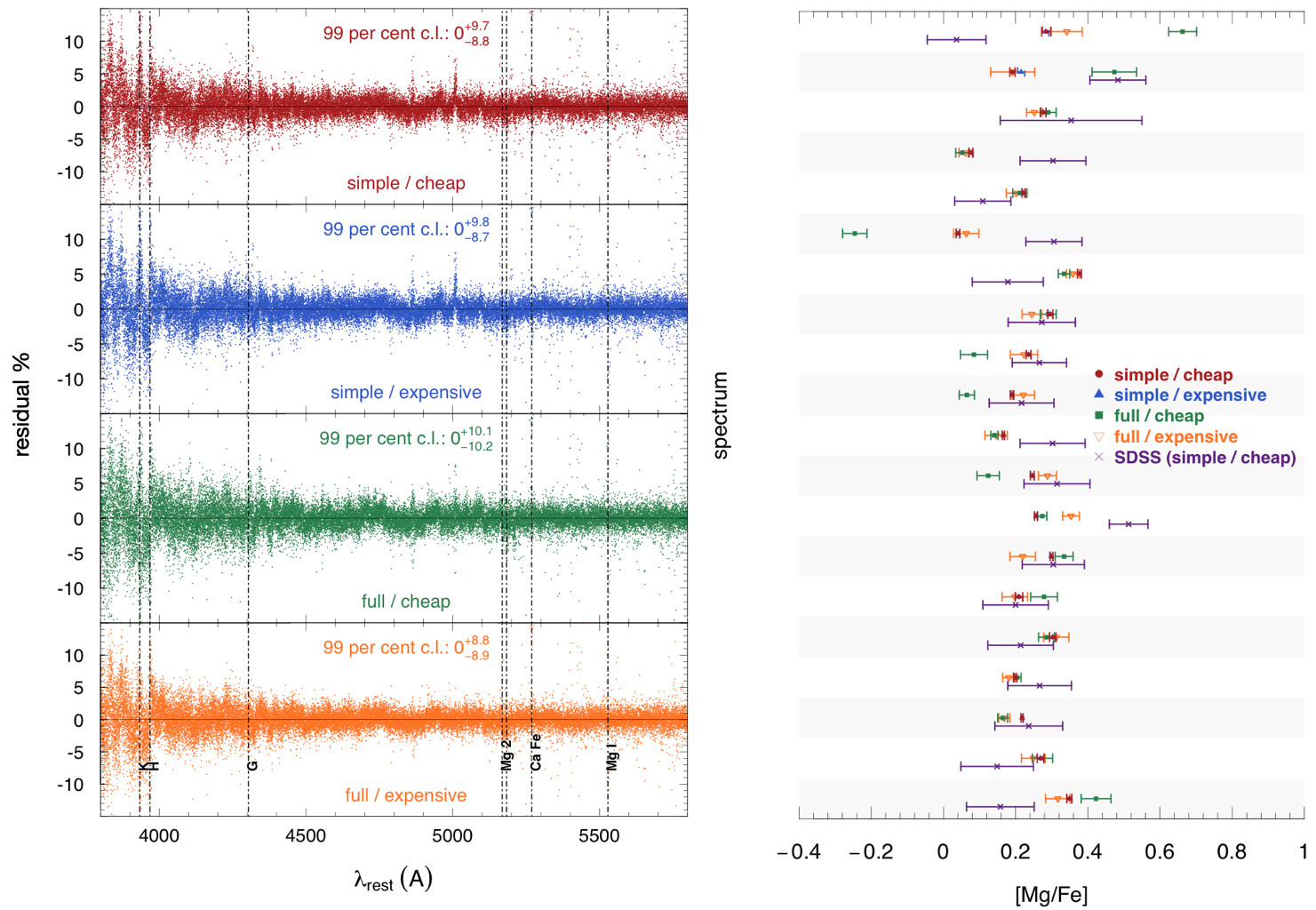} \caption{{\it Left}: Fractional residual, i.e., 100 $\times$ (1 - [model / data]) of all the 20 spectra considered for  {\tt alf}'s performance test in  four different setups. The main absorption features are shown in dashed lines. {\it Right}: [Mg/Fe] obtained through each setup. Each white/gray strip corresponds to one spectrum. For a comparison, slightly offset in the vertical direction from these measurements, we show the [Mg/Fe] measured from the SDSS spectra (that is, from single fiber as opposed to the integrated MaNGA spectra) using the simple/cheap setup.}
\label{fig:performance}
\end{center}
\end{figure}

All setups resulted in excellent synthetic spectra, as demonstrated in Fig.\ref{fig:performance}. The quality is evident from the negligible fractional residuals and their low spread ($0\pm10\%$ at 99\% c.~l.). {\tt alf} also performs well in parameter estimation, with good agreement observed among the [Mg/Fe] abundances obtained in each configuration (right panel of Fig.\ref{fig:performance}). The only notable exception is the parameters obtained in full mode with the cheap MCMC configuration, likely due to the inability of the MCMC chains to stabilize. Consequently, considering the shorter processing time per galaxy and the smaller error bars in the final parameters, we chose the simple mode under the cheap MCMC configuration for our fiducial measurements.

We also compared the [Mg/Fe] estimated from our integrated MaNGA spectra with those from the SDSS. As can be seen in Fig.~\ref{fig:performance} (right panel), the high quality of the MaNGA data resulted in much smaller error bars when compared with the SDSS spectra. The average error in [Mg/Fe] was respectively 0.006 and 0.092 in our testing sample.

\begin{figure}
\begin{center}
\includegraphics[width=1.0\textwidth]{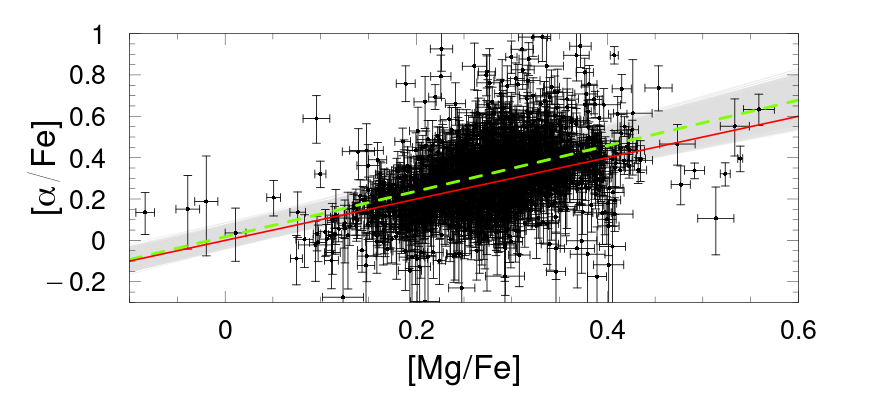} \caption{Comparison between the [Mg/Fe] measurements obtained in this paper and the [$\alpha$/Fe] estimates from \cite{Gallazzi21}. The dashed green line represents the resulting linear regression that accounts for the error bars on both axes. This relation is comparable to the curve $x=y$ (red continuous line) within a 68\% c.~l. (gray straight lines), a confidence region obtained after $10^4$ re-samplings of the linear regression coefficients.}
\label{fig:galazzi}
\end{center}
\end{figure}

To complete our assessment of {\rm alf} measurements, we compared its outcome with the [$\alpha$/Fe] abundance catalog determined by \cite{Gallazzi21}. The measurements of 1605 galaxies found in common with our MaNGA sample are illustrated in Fig.~\ref{fig:galazzi}. The resulting linear regression \citep[{\sc R} package {\sc deming};][]{deming}, presented in Fig.~\ref{fig:galazzi}, closely aligns with the one-to-one curve within a 68\% confidence level. The considerably larger error bars in \cite{Gallazzi21}'s sample are most likely due to the fact that they used SDSS single-fiber  spectra from DR7.

\section{{\tt ellipse} performance test}
\label{sec:ellipse.test}

In order to validate the robustness of our isophotal shape classification, anchored in the parameter $a_4/a$ as defined by Eq.~\ref{eq.a4}, we meticulously examined the accuracy with which the \texttt{ellipse} tool retrieves both the fourth cosine coefficient ($B_4$) and the ellipticity ($\epsilon$). To accomplish this, we generated a mock sample comprising 200 EGs (300 $\times$ 300 pixels), evenly divided between pristine (noiseless) images and those injected with noise to emulate realistic observational conditions. The median value of the mock  $a_4/a$ and its corresponding 68\% c.~l. were  0.0013 and [-0.042, 0.093], respectively, for both the noise-free and realistic datasets. Subsequently, we applied the identical measurement pipeline introduced in Section~\ref{sec:light} to recover $a_4/a$. This calculation was performed within the semi-major axis interval of  $5 \leq a \leq 30$ pixels, aiming to exclude both the innermost part affected by the seeing effect and the low S/N peripheral regions in real images.

\begin{figure}
\begin{center}
\includegraphics[width=1.0\textwidth]{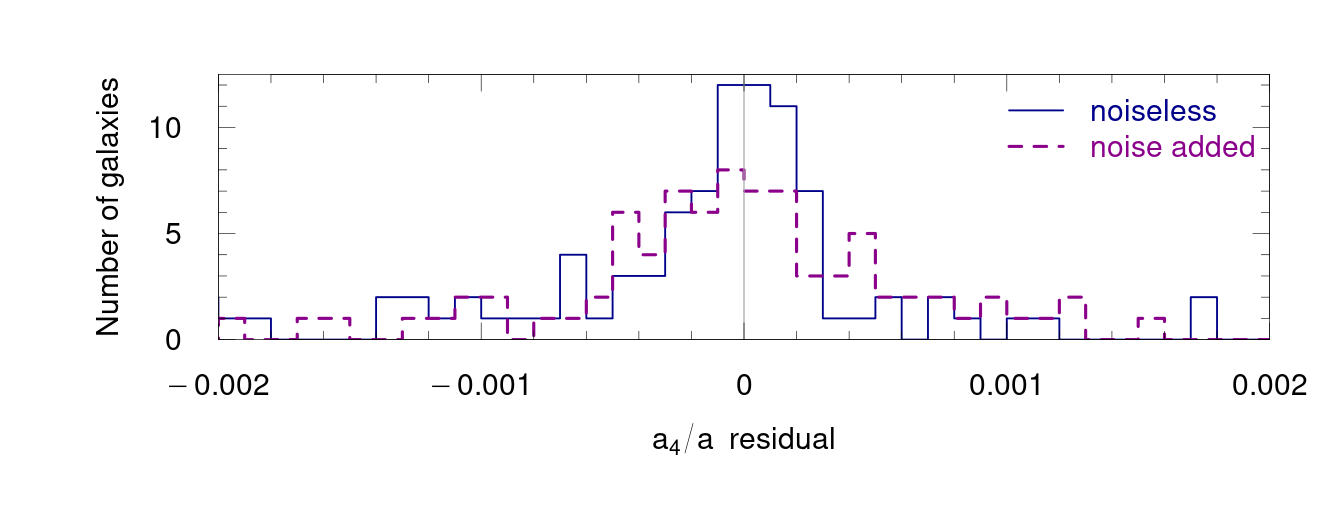} \caption{Histogram of the residuals (mock - {\tt ellipse}) in $a_4/a$. This is derived from two sets of 100 mock elliptical galaxies: one set consists of noiseless images, and the other set mimics realistic images. Both samples exhibit a null median residual with a very small dispersion.}
\label{fig:ellipse.performance}
\end{center}
\end{figure}

Our findings, presented in Fig.~\ref{fig:ellipse.performance}, confirm the high fidelity of the \texttt{ellipse} package in accurately reconstructing the isophotal shape, as evidenced by the sharp residuals centered around zero. The 68\% c.~l.~ranges within the intervals [$-0.0008:0.0004$] and [$-0.0005:0.0011$], respectively, for noiseless and realistic mock images. This indicates a reliable classification in terms of boxiness or diskiness, with only one noisy image (out of 100) exhibiting a classification different from the truth.

\section{Higher order  principal components (PC3 to PC7)}
\label{sec:pca.higher}

\begin{figure}
\begin{center}
\includegraphics[width=1.0\textwidth]{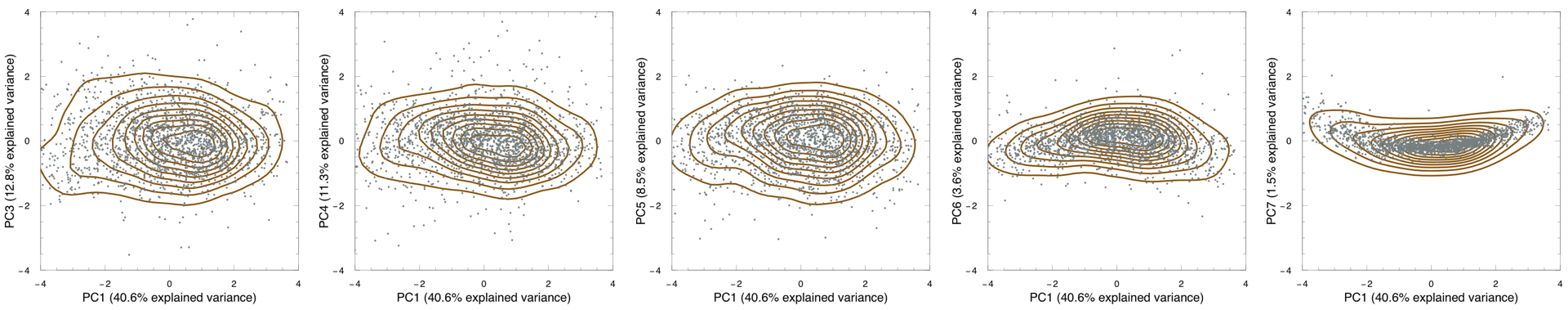} \caption{Comparison of the first principal component (PC1) with the 3rd, 4th, 5th, 6th and 7th  principal components.}
\label{fig:high.pca}
\end{center}
\end{figure}

For the sake of completeness, we present in Fig.~\ref{fig:high.pca} the higher order  principal components, PC3, PC4, PC5, PC6,  and PC7. Starting from PC1 and summing up until the 7th order, they account for 95\% of the total variance in the MaNGA sample. Consistent with the conclusions drawn from the first two PCs, no clustering in these parameter space is observed.

\section{Correlation among the parameters}
\label{sec:corr}

\begin{figure}
\begin{center}
\includegraphics[width=0.6\textwidth]{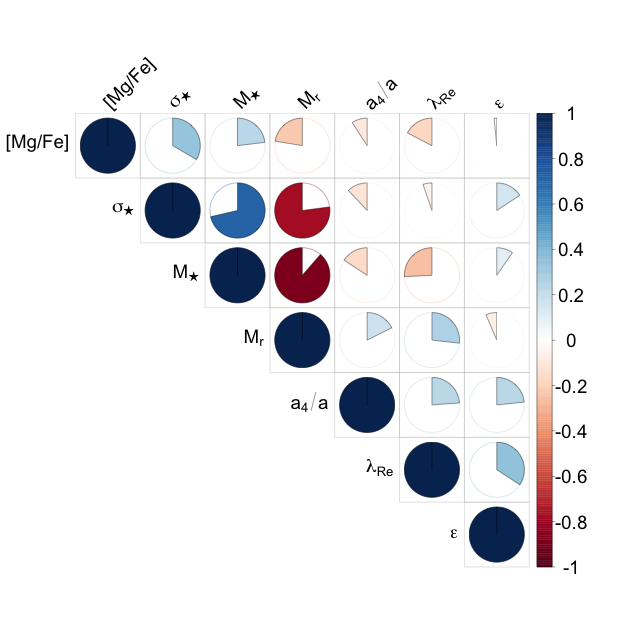} \caption{Correlation among the physical parameters in this work, measured across our MaNGA EG sample. The magnitude of the correlation is represented both by the area of the circles and the color scale, with fully filled dark blue circles indicating a perfect positive correlation (correlation coefficient of 1) and fully filled dark red circles representing a perfect negative correlation (correlation coefficient of $-1$).}
\label{fig:corr}
\end{center}
\end{figure}

In Fig.~\ref{fig:corr} we present the correlation among all the seven parameters considered in this work, as listed in Eq.~\ref{eq:param.vector}. Overall, the correlations and anti-correlations are negligible (i.e., with absolute values less than 0.5), except for those involving absolute magnitude, stellar mass, and velocity dispersion, which exhibit stronger (anti) correlations among them due to their intrinsic physical connections.

\clearpage

\section{Measured parameters}
\label{sec:parameters}

In this work, we present new measurements of the parameters for EGs, including the isophotal shape parameter (Section~\ref{sec:light}), the luminosity-weighted stellar angular momentum parameter (Section.~\ref{sec:dynamics}), the stellar velocity dispersion (Section~\ref{sec:chemical}), and the magnesium-to-iron ratio (Section~\ref{sec:chemical}), with their nominal values and corresponding uncertainties provided in Tab.~\ref{tab:parameters.full}. The RA and DEC were retrieved directly from NSA (Section~\ref{sec:nsa}).

\begin{center}
\begin{longtable}{c c c c c c}
\caption{EGs parameters. The complete version is available in the online Journal in the machine-readable format.} \label{tab:parameters.full} \\

\hline \hline
\multicolumn{1}{c}{RA (deg)} & \multicolumn{1}{c}{DEC (deg)} & \multicolumn{1}{c}{$\langle a_4/a \rangle$}  & \multicolumn{1}{c}{$\lambda_{\rm Re}$} & \multicolumn{1}{c}{$\sigma_\star$ (km s$^{-1}$)} & \multicolumn{1}{c}{$[{\rm Mg/Fe}]$} \\ \hline \hline
\endfirsthead

\multicolumn{6}{c}
{{\bfseries \tablename\ \thetable{} -- continued from previous page}} \\
\hline 
\multicolumn{1}{c}{RA (deg)} & \multicolumn{1}{c}{DEC (deg)} & \multicolumn{1}{c}{$\langle a_4/a \rangle$}  & \multicolumn{1}{c}{$\lambda_{\rm Re}$} & \multicolumn{1}{c}{$\sigma_\star$ (km s$^{-1}$)} & \multicolumn{1}{c}{$[{\rm Mg/Fe}]$} \\ \hline 
\endhead

\hline \multicolumn{6}{|r|}{{Continued on next page}} \\ \hline
\endfoot

\hline \hline
\endlastfoot

  182.3563 &     46.5494 & $-0.0016\pm 0.0011$ & $  0.149_{ -0.111}^{+  0.007}$ & $163.1\pm 1.0$ & $  0.250\pm  0.008$ \\
  117.3978 &     30.4166 & $-0.0021\pm 0.0007$ & $  0.056_{ -0.019}^{+  0.002}$ & $215.6\pm 0.7$ & $  0.309\pm  0.005$ \\
  203.7853 &     42.6472 & $-0.0054\pm 0.0008$ & $  0.081_{ -0.047}^{+  0.003}$ & $183.6\pm 0.8$ & $  0.350\pm  0.006$ \\
  115.3875 &     47.8710 & $-0.0072\pm 0.0012$ & $  0.168_{ -0.072}^{+  0.005}$ & $212.6\pm 2.3$ & $  0.274\pm  0.017$ \\
  226.2318 &     31.1250 & $ 0.0072\pm 0.0005$ & $  0.078_{ -0.040}^{+  0.002}$ & $212.2\pm 0.7$ & $  0.221\pm  0.005$ \\
  \ldots   & \ldots      & \ldots              &  \ldots                        &  \ldots        &  \ldots             \\
  355.9464 &      0.7059 & $-0.0046\pm 0.0005$ & $  0.198_{ -0.168}^{+  0.010}$ & $220.1\pm 1.1$ & $  0.293\pm  0.010$ \\
  355.0894 &      0.2695 & $-0.0025\pm 0.0006$ & $  0.186_{ -0.129}^{+  0.008}$ & $238.5\pm 2.3$ & $  0.326\pm  0.013$ \\
  355.2802 &      0.0952 & $ 0.0040\pm 0.0003$ & $  0.056_{ -0.066}^{+  0.004}$ & $244.8\pm 0.8$ & $  0.345\pm  0.005$ \\
  355.1005 &      0.0023 & $ 0.0003\pm 0.0004$ & $  0.228_{ -0.110}^{+  0.007}$ & $168.0\pm 1.0$ & $  0.295\pm  0.008$ \\
  359.8362 &     14.6702 & $-0.0141\pm 0.0019$ & $  0.092_{ -0.093}^{+  0.006}$ & $196.5\pm 1.2$ & $  0.315\pm  0.010$ \\

\end{longtable}
\end{center}

\clearpage\newpage
\bibliography{paper}{}

\begin{thebibliography}{}
\expandafter\ifx\csname natexlab\endcsname\relax\def\natexlab#1{#1}\fi
\providecommand{\url}[1]{\href{#1}{#1}}
\providecommand{\dodoi}[1]{doi:~\href{http://doi.org/#1}{\nolinkurl{#1}}}
\providecommand{\doeprint}[1]{\href{http://ascl.net/#1}{\nolinkurl{http://ascl.net/#1}}}
\providecommand{\doarXiv}[1]{\href{https://arxiv.org/abs/#1}{\nolinkurl{https://arxiv.org/abs/#1}}}

\bibitem[{{Abazajian} {et~al.}(2009){Abazajian}, {Adelman-McCarthy}, {Ag{\"u}eros}, {Allam}, {Allende Prieto}, {An}, {Anderson}, {Anderson}, {Annis}, {Bahcall}, {Bailer-Jones}, {Barentine}, {Bassett}, {Becker}, {Beers}, {Bell}, {Belokurov}, {Berlind}, {Berman}, {Bernardi}, {Bickerton}, {Bizyaev}, {Blakeslee}, {Blanton}, {Bochanski}, {Boroski}, {Brewington}, {Brinchmann}, {Brinkmann}, {Brunner}, {Budav{\'a}ri}, {Carey}, {Carliles}, {Carr}, {Castander}, {Cinabro}, {Connolly}, {Csabai}, {Cunha}, {Czarapata}, {Davenport}, {de Haas}, {Dilday}, {Doi}, {Eisenstein}, {Evans}, {Evans}, {Fan}, {Friedman}, {Frieman}, {Fukugita}, {G{\"a}nsicke}, {Gates}, {Gillespie}, {Gilmore}, {Gonzalez}, {Gonzalez}, {Grebel}, {Gunn}, {Gy{\"o}ry}, {Hall}, {Harding}, {Harris}, {Harvanek}, {Hawley}, {Hayes}, {Heckman}, {Hendry}, {Hennessy}, {Hindsley}, {Hoblitt}, {Hogan}, {Hogg}, {Holtzman}, {Hyde}, {Ichikawa}, {Ichikawa}, {Im}, {Ivezi{\'c}}, {Jester}, {Jiang}, {Johnson}, {Jorgensen}, {Juri{\'c}}, {Kent}, {Kessler}, {Kleinman}, {Knapp},
  {Konishi}, {Kron}, {Krzesinski}, {Kuropatkin}, {Lampeitl}, {Lebedeva}, {Lee}, {Lee}, {French Leger}, {L{\'e}pine}, {Li}, {Lima}, {Lin}, {Long}, {Loomis}, {Loveday}, {Lupton}, {Magnier}, {Malanushenko}, {Malanushenko}, {Mandelbaum}, {Margon}, {Marriner}, {Mart{\'\i}nez-Delgado}, {Matsubara}, {McGehee}, {McKay}, {Meiksin}, {Morrison}, {Mullally}, {Munn}, {Murphy}, {Nash}, {Nebot}, {Neilsen}, {Newberg}, {Newman}, {Nichol}, {Nicinski}, {Nieto-Santisteban}, {Nitta}, {Okamura}, {Oravetz}, {Ostriker}, {Owen}, {Padmanabhan}, {Pan}, {Park}, {Pauls}, {Peoples}, {Percival}, {Pier}, {Pope}, {Pourbaix}, {Price}, {Purger}, {Quinn}, {Raddick}, {Re Fiorentin}, {Richards}, {Richmond}, {Riess}, {Rix}, {Rockosi}, {Sako}, {Schlegel}, {Schneider}, {Scholz}, {Schreiber}, {Schwope}, {Seljak}, {Sesar}, {Sheldon}, {Shimasaku}, {Sibley}, {Simmons}, {Sivarani}, {Allyn Smith}, {Smith}, {Smol{\v{c}}i{\'c}}, {Snedden}, {Stebbins}, {Steinmetz}, {Stoughton}, {Strauss}, {SubbaRao}, {Suto}, {Szalay}, {Szapudi}, {Szkody}, {Tanaka},
  {Tegmark}, {Teodoro}, {Thakar}, {Tremonti}, {Tucker}, {Uomoto}, {Vanden Berk}, {Vandenberg}, {Vidrih}, {Vogeley}, {Voges}, {Vogt}, {Wadadekar}, {Watters}, {Weinberg}, {West}, {White}, {Wilhite}, {Wonders}, {Yanny}, {Yocum}, {York}, {Zehavi}, {Zibetti}, \& {Zucker}}]{SDSS-DR7}
{Abazajian}, K.~N., {Adelman-McCarthy}, J.~K., {Ag{\"u}eros}, M.~A., {et~al.} 2009, \apjs, 182, 543, \dodoi{10.1088/0067-0049/182/2/543}

\bibitem[{{Abdurro'uf} {et~al.}(2022){Abdurro'uf}, {Accetta}, {Aerts}, {Silva Aguirre}, {Ahumada}, {Ajgaonkar}, {Filiz Ak}, {Alam}, {Allende Prieto}, {Almeida}, {Anders}, {Anderson}, {Andrews}, {Anguiano}, {Aquino-Ort{\'\i}z}, {Arag{\'o}n-Salamanca}, {Argudo-Fern{\'a}ndez}, {Ata}, {Aubert}, {Avila-Reese}, {Badenes}, {Barb{\'a}}, {Barger}, {Barrera-Ballesteros}, {Beaton}, {Beers}, {Belfiore}, {Bender}, {Bernardi}, {Bershady}, {Beutler}, {Bidin}, {Bird}, {Bizyaev}, {Blanc}, {Blanton}, {Boardman}, {Bolton}, {Boquien}, {Borissova}, {Bovy}, {Brandt}, {Brown}, {Brownstein}, {Brusa}, {Buchner}, {Bundy}, {Burchett}, {Bureau}, {Burgasser}, {Cabang}, {Campbell}, {Cappellari}, {Carlberg}, {Wanderley}, {Carrera}, {Cash}, {Chen}, {Chen}, {Cherinka}, {Chiappini}, {Choi}, {Chojnowski}, {Chung}, {Clerc}, {Cohen}, {Comerford}, {Comparat}, {da Costa}, {Covey}, {Crane}, {Cruz-Gonzalez}, {Culhane}, {Cunha}, {Dai}, {Damke}, {Darling}, {Davidson}, {Davies}, {Dawson}, {De Lee}, {Diamond-Stanic}, {Cano-D{\'\i}az}, {S{\'a}nchez},
  {Donor}, {Duckworth}, {Dwelly}, {Eisenstein}, {Elsworth}, {Emsellem}, {Eracleous}, {Escoffier}, {Fan}, {Farr}, {Feng}, {Fern{\'a}ndez-Trincado}, {Feuillet}, {Filipp}, {Fillingham}, {Frinchaboy}, {Fromenteau}, {Galbany}, {Garc{\'\i}a}, {Garc{\'\i}a-Hern{\'a}ndez}, {Ge}, {Geisler}, {Gelfand}, {G{\'e}ron}, {Gibson}, {Goddy}, {Godoy-Rivera}, {Grabowski}, {Green}, {Greener}, {Grier}, {Griffith}, {Guo}, {Guy}, {Hadjara}, {Harding}, {Hasselquist}, {Hayes}, {Hearty}, {Hern{\'a}ndez}, {Hill}, {Hogg}, {Holtzman}, {Horta}, {Hsieh}, {Hsu}, {Hsu}, {Huber}, {Huertas-Company}, {Hutchinson}, {Hwang}, {Ibarra-Medel}, {Chitham}, {Ilha}, {Imig}, {Jaekle}, {Jayasinghe}, {Ji}, {Johnson}, {Jones}, {J{\"o}nsson}, {Katkov}, {Khalatyan}, {Kinemuchi}, {Kisku}, {Knapen}, {Kneib}, {Kollmeier}, {Kong}, {Kounkel}, {Kreckel}, {Krishnarao}, {Lacerna}, {Lane}, {Langgin}, {Lavender}, {Law}, {Lazarz}, {Leung}, {Leung}, {Lewis}, {Li}, {Li}, {Lian}, {Liang}, {Lin}, {Lin}, {Lin}, {Lintott}, {Long}, {Longa-Pe{\~n}a}, {L{\'o}pez-Cob{\'a}}, {Lu},
  {Lundgren}, {Luo}, {Mackereth}, {de la Macorra}, {Mahadevan}, {Majewski}, {Manchado}, {Mandeville}, {Maraston}, {Margalef-Bentabol}, {Masseron}, {Masters}, {Mathur}, {McDermid}, {Mckay}, {Merloni}, {Merrifield}, {Meszaros}, {Miglio}, {Di Mille}, {Minniti}, {Minsley}, {Monachesi}, {Moon}, {Mosser}, {Mulchaey}, {Muna}, {Mu{\~n}oz}, {Myers}, {Myers}, {Nadathur}, {Nair}, {Nandra}, {Neumann}, {Newman}, {Nidever}, {Nikakhtar}, {Nitschelm}, {O'Connell}, {Garma-Oehmichen}, {Luan Souza de Oliveira}, {Olney}, {Oravetz}, {Ortigoza-Urdaneta}, {Osorio}, {Otter}, {Pace}, {Padilla}, {Pan}, {Pan}, {Parikh}, {Parker}, {Peirani}, {Pe{\~n}a Ram{\'\i}rez}, {Penny}, {Percival}, {Perez-Fournon}, {Pinsonneault}, {Poidevin}, {Poovelil}, {Price-Whelan}, {B{\'a}rbara de Andrade Queiroz}, {Raddick}, {Ray}, {Rembold}, {Riddle}, {Riffel}, {Riffel}, {Rix}, {Robin}, {Rodr{\'\i}guez-Puebla}, {Roman-Lopes}, {Rom{\'a}n-Z{\'u}{\~n}iga}, {Rose}, {Ross}, {Rossi}, {Rubin}, {Salvato}, {S{\'a}nchez}, {S{\'a}nchez-Gallego}, {Sanderson}, {Santana
  Rojas}, {Sarceno}, {Sarmiento}, {Sayres}, {Sazonova}, {Schaefer}, {Schiavon}, {Schlegel}, {Schneider}, {Schultheis}, {Schwope}, {Serenelli}, {Serna}, {Shao}, {Shapiro}, {Sharma}, {Shen}, {Shetrone}, {Shu}, {Simon}, {Skrutskie}, {Smethurst}, {Smith}, {Sobeck}, {Spoo}, {Sprague}, {Stark}, {Stassun}, {Steinmetz}, {Stello}, {Stone-Martinez}, {Storchi-Bergmann}, {Stringfellow}, {Stutz}, {Su}, {Taghizadeh-Popp}, {Talbot}, {Tayar}, {Telles}, {Teske}, {Thakar}, {Theissen}, {Tkachenko}, {Thomas}, {Tojeiro}, {Hernandez Toledo}, {Troup}, {Trump}, {Trussler}, {Turner}, {Tuttle}, {Unda-Sanzana}, {V{\'a}zquez-Mata}, {Valentini}, {Valenzuela}, {Vargas-Gonz{\'a}lez}, {Vargas-Maga{\~n}a}, {Alfaro}, {Villanova}, {Vincenzo}, {Wake}, {Warfield}, {Washington}, {Weaver}, {Weijmans}, {Weinberg}, {Weiss}, {Westfall}, {Wild}, {Wilde}, {Wilson}, {Wilson}, {Wilson}, {Wolf}, {Wood-Vasey}, {Yan}, {Zamora}, {Zasowski}, {Zhang}, {Zhao}, {Zheng}, {Zheng}, \& {Zhu}}]{Abdurrouf22}
{Abdurro'uf}, {Accetta}, K., {Aerts}, C., {et~al.} 2022, \apjs, 259, 35, \dodoi{10.3847/1538-4365/ac4414}

\bibitem[{{Adelman-McCarthy} {et~al.}(2006){Adelman-McCarthy}, {Ag{\"u}eros}, {Allam}, {Anderson}, {Anderson}, {Annis}, {Bahcall}, {Baldry}, {Barentine}, {Berlind}, {Bernardi}, {Blanton}, {Boroski}, {Brewington}, {Brinchmann}, {Brinkmann}, {Brunner}, {Budav{\'a}ri}, {Carey}, {Carr}, {Castander}, {Connolly}, {Csabai}, {Czarapata}, {Dalcanton}, {Doi}, {Dong}, {Eisenstein}, {Evans}, {Fan}, {Finkbeiner}, {Friedman}, {Frieman}, {Fukugita}, {Gillespie}, {Glazebrook}, {Gray}, {Grebel}, {Gunn}, {Gurbani}, {de Haas}, {Hall}, {Harris}, {Harvanek}, {Hawley}, {Hayes}, {Hendry}, {Hennessy}, {Hindsley}, {Hirata}, {Hogan}, {Hogg}, {Holmgren}, {Holtzman}, {Ichikawa}, {Ivezi{\'c}}, {Jester}, {Johnston}, {Jorgensen}, {Juri{\'c}}, {Kent}, {Kleinman}, {Knapp}, {Kniazev}, {Kron}, {Krzesinski}, {Kuropatkin}, {Lamb}, {Lampeitl}, {Lee}, {Leger}, {Lin}, {Long}, {Loveday}, {Lupton}, {Margon}, {Mart{\'\i}nez-Delgado}, {Mandelbaum}, {Matsubara}, {McGehee}, {McKay}, {Meiksin}, {Munn}, {Nakajima}, {Nash}, {Neilsen}, {Newberg}, {Newman},
  {Nichol}, {Nicinski}, {Nieto-Santisteban}, {Nitta}, {O'Mullane}, {Okamura}, {Owen}, {Padmanabhan}, {Pauls}, {Peoples}, {Pier}, {Pope}, {Pourbaix}, {Quinn}, {Richards}, {Richmond}, {Rockosi}, {Schlegel}, {Schneider}, {Schroeder}, {Scranton}, {Seljak}, {Sheldon}, {Shimasaku}, {Smith}, {Smol{\v{c}}i{\'c}}, {Snedden}, {Stoughton}, {Strauss}, {SubbaRao}, {Szalay}, {Szapudi}, {Szkody}, {Tegmark}, {Thakar}, {Tucker}, {Uomoto}, {Vanden Berk}, {Vandenberg}, {Vogeley}, {Voges}, {Vogt}, {Walkowicz}, {Weinberg}, {West}, {White}, {Xu}, {Yanny}, {Yocum}, {York}, {Zehavi}, {Zibetti}, \& {Zucker}}]{SDSS-DR4}
{Adelman-McCarthy}, J.~K., {Ag{\"u}eros}, M.~A., {Allam}, S.~S., {et~al.} 2006, \apjs, 162, 38, \dodoi{10.1086/497917}

\bibitem[{{Albuquerque} {et~al.}(2024){Albuquerque}, {Machado}, \& {Monteiro-Oliveira}}]{Albuquerque24}
{Albuquerque}, R.~P., {Machado}, R. E.~G., \& {Monteiro-Oliveira}, R. 2024, arXiv e-prints, arXiv:2401.15044, \dodoi{10.48550/arXiv.2401.15044}

\bibitem[{{Bender} {et~al.}(1988){Bender}, {Doebereiner}, \& {Moellenhoff}}]{Bender88}
{Bender}, R., {Doebereiner}, S., \& {Moellenhoff}, C. 1988, \aaps, 74, 385

\bibitem[{{Bender} {et~al.}(1989){Bender}, {Surma}, {Doebereiner}, {Moellenhoff}, \& {Madejsky}}]{Bender89}
{Bender}, R., {Surma}, P., {Doebereiner}, S., {Moellenhoff}, C., \& {Madejsky}, R. 1989, \aap, 217, 35

\bibitem[{{Bernardi} {et~al.}(2023){Bernardi}, {Dom{\'\i}nguez S{\'a}nchez}, {Sheth}, {Brownstein}, \& {Lane}}]{Bernardi23}
{Bernardi}, M., {Dom{\'\i}nguez S{\'a}nchez}, H., {Sheth}, R.~K., {Brownstein}, J.~R., \& {Lane}, R.~R. 2023, \mnras, 518, 4713, \dodoi{10.1093/mnras/stac3287}

\bibitem[{{Bertin} \& {Arnouts}(1996)}]{sextractor}
{Bertin}, E., \& {Arnouts}, S. 1996, \aaps, 117, 393

\bibitem[{{Best} \& {Heckman}(2012)}]{Best12}
{Best}, P.~N., \& {Heckman}, T.~M. 2012, \mnras, 421, 1569, \dodoi{10.1111/j.1365-2966.2012.20414.x}

\bibitem[{{Best} {et~al.}(2005){Best}, {Kauffmann}, {Heckman}, {Brinchmann}, {Charlot}, {Ivezi{\'c}}, \& {White}}]{Best05}
{Best}, P.~N., {Kauffmann}, G., {Heckman}, T.~M., {et~al.} 2005, \mnras, 362, 25, \dodoi{10.1111/j.1365-2966.2005.09192.x}

\bibitem[{{B{\'\i}lek} {et~al.}(2023){B{\'\i}lek}, {Duc}, \& {Sola}}]{Bilek23}
{B{\'\i}lek}, M., {Duc}, P.~A., \& {Sola}, E. 2023, \aap, 672, A27, \dodoi{10.1051/0004-6361/202244749}

\bibitem[{{Binggeli} \& {Cameron}(1991)}]{Binggeli91}
{Binggeli}, B., \& {Cameron}, L.~M. 1991, \aap, 252, 27

\bibitem[{{Binney} \& {Tremaine}(2008)}]{Binney08}
{Binney}, J., \& {Tremaine}, S. 2008, {Galactic Dynamics: Second Edition}

\bibitem[{{Blanton} {et~al.}(2011){Blanton}, {Kazin}, {Muna}, {Weaver}, \& {Price-Whelan}}]{Blanton11}
{Blanton}, M.~R., {Kazin}, E., {Muna}, D., {Weaver}, B.~A., \& {Price-Whelan}, A. 2011, \aj, 142, 31, \dodoi{10.1088/0004-6256/142/1/31}

\bibitem[{{Bois} {et~al.}(2011){Bois}, {Emsellem}, {Bournaud}, {Alatalo}, {Blitz}, {Bureau}, {Cappellari}, {Davies}, {Davis}, {de Zeeuw}, {Duc}, {Khochfar}, {Krajnovi{\'c}}, {Kuntschner}, {Lablanche}, {McDermid}, {Morganti}, {Naab}, {Oosterloo}, {Sarzi}, {Scott}, {Serra}, {Weijmans}, \& {Young}}]{Bois11}
{Bois}, M., {Emsellem}, E., {Bournaud}, F., {et~al.} 2011, \mnras, 416, 1654, \dodoi{10.1111/j.1365-2966.2011.19113.x}

\bibitem[{{Bundy} {et~al.}(2015){Bundy}, {Bershady}, {Law}, {Yan}, {Drory}, {MacDonald}, {Wake}, {Cherinka}, {S{\'a}nchez-Gallego}, {Weijmans}, {Thomas}, {Tremonti}, {Masters}, {Coccato}, {Diamond-Stanic}, {Arag{\'o}n-Salamanca}, {Avila-Reese}, {Badenes}, {Falc{\'o}n-Barroso}, {Belfiore}, {Bizyaev}, {Blanc}, {Bland-Hawthorn}, {Blanton}, {Brownstein}, {Byler}, {Cappellari}, {Conroy}, {Dutton}, {Emsellem}, {Etherington}, {Frinchaboy}, {Fu}, {Gunn}, {Harding}, {Johnston}, {Kauffmann}, {Kinemuchi}, {Klaene}, {Knapen}, {Leauthaud}, {Li}, {Lin}, {Maiolino}, {Malanushenko}, {Malanushenko}, {Mao}, {Maraston}, {McDermid}, {Merrifield}, {Nichol}, {Oravetz}, {Pan}, {Parejko}, {Sanchez}, {Schlegel}, {Simmons}, {Steele}, {Steinmetz}, {Thanjavur}, {Thompson}, {Tinker}, {van den Bosch}, {Westfall}, {Wilkinson}, {Wright}, {Xiao}, \& {Zhang}}]{Bundy15}
{Bundy}, K., {Bershady}, M.~A., {Law}, D.~R., {et~al.} 2015, \apj, 798, 7, \dodoi{10.1088/0004-637X/798/1/7}

\bibitem[{{Cappellari}(2016)}]{Cappellari16}
{Cappellari}, M. 2016, \araa, 54, 597, \dodoi{10.1146/annurev-astro-082214-122432}

\bibitem[{{Cappellari} {et~al.}(2011){Cappellari}, {Emsellem}, {Krajnovi{\'c}}, {McDermid}, {Serra}, {Alatalo}, {Blitz}, {Bois}, {Bournaud}, {Bureau}, {Davies}, {Davis}, {de Zeeuw}, {Khochfar}, {Kuntschner}, {Lablanche}, {Morganti}, {Naab}, {Oosterloo}, {Sarzi}, {Scott}, {Weijmans}, \& {Young}}]{Cappellari11}
{Cappellari}, M., {Emsellem}, E., {Krajnovi{\'c}}, D., {et~al.} 2011, \mnras, 416, 1680, \dodoi{10.1111/j.1365-2966.2011.18600.x}

\bibitem[{{Chaware} {et~al.}(2014){Chaware}, {Cannon}, {Kembhavi}, {Mahabal}, \& {Pandey}}]{Chaware14}
{Chaware}, L., {Cannon}, R., {Kembhavi}, A.~K., {Mahabal}, A., \& {Pandey}, S.~K. 2014, \apj, 787, 102, \dodoi{10.1088/0004-637X/787/2/102}

\bibitem[{{Choi} {et~al.}(2016){Choi}, {Dotter}, {Conroy}, {Cantiello}, {Paxton}, \& {Johnson}}]{Choi16}
{Choi}, J., {Dotter}, A., {Conroy}, C., {et~al.} 2016, \apj, 823, 102, \dodoi{10.3847/0004-637X/823/2/102}

\bibitem[{{Ciambur}(2015)}]{Ciambur15}
{Ciambur}, B.~C. 2015, \apj, 810, 120, \dodoi{10.1088/0004-637X/810/2/120}

\bibitem[{{Conroy} {et~al.}(2014){Conroy}, {Graves}, \& {van Dokkum}}]{Conroy14}
{Conroy}, C., {Graves}, G.~J., \& {van Dokkum}, P.~G. 2014, \apj, 780, 33, \dodoi{10.1088/0004-637X/780/1/33}

\bibitem[{{Conroy} \& {van Dokkum}(2012)}]{Conroy12}
{Conroy}, C., \& {van Dokkum}, P. 2012, \apj, 747, 69, \dodoi{10.1088/0004-637X/747/1/69}

\bibitem[{{Conroy} {et~al.}(2018){Conroy}, {Villaume}, {van Dokkum}, \& {Lind}}]{Conroy18}
{Conroy}, C., {Villaume}, A., {van Dokkum}, P.~G., \& {Lind}, K. 2018, \apj, 854, 139, \dodoi{10.3847/1538-4357/aaab49}

\bibitem[{{C{\^o}t{\'e}} {et~al.}(2006){C{\^o}t{\'e}}, {Piatek}, {Ferrarese}, {Jord{\'a}n}, {Merritt}, {Peng}, {Ha{\c{s}}egan}, {Blakeslee}, {Mei}, {West}, {Milosavljevi{\'c}}, \& {Tonry}}]{Cote06}
{C{\^o}t{\'e}}, P., {Piatek}, S., {Ferrarese}, L., {et~al.} 2006, \apjs, 165, 57, \dodoi{10.1086/504042}

\bibitem[{{Dalal} {et~al.}(2021){Dalal}, {Strauss}, {Sunayama}, {Oguri}, {Lin}, {Huang}, {Park}, \& {Takada}}]{Dalal21}
{Dalal}, R., {Strauss}, M.~A., {Sunayama}, T., {et~al.} 2021, \mnras, 507, 4016, \dodoi{10.1093/mnras/stab2363}

\bibitem[{{Davies} {et~al.}(1983){Davies}, {Efstathiou}, {Fall}, {Illingworth}, \& {Schechter}}]{Davies83}
{Davies}, R.~L., {Efstathiou}, G., {Fall}, S.~M., {Illingworth}, G., \& {Schechter}, P.~L. 1983, \apj, 266, 41, \dodoi{10.1086/160757}

\bibitem[{{Ding} {et~al.}(2025){Ding}, {Dalal}, {Sunayama}, {Strauss}, {Oguri}, {Okabe}, {Hilton}, {Monteiro-Oliveira}, {Sif{\'o}n}, \& {Staggs}}]{Ding25}
{Ding}, J., {Dalal}, R., {Sunayama}, T., {et~al.} 2025, \mnras, 536, 572, \dodoi{10.1093/mnras/stae2601}

\bibitem[{{Dom{\'\i}nguez S{\'a}nchez} {et~al.}(2018){Dom{\'\i}nguez S{\'a}nchez}, {Huertas-Company}, {Bernardi}, {Tuccillo}, \& {Fischer}}]{DSanchez18}
{Dom{\'\i}nguez S{\'a}nchez}, H., {Huertas-Company}, M., {Bernardi}, M., {Tuccillo}, D., \& {Fischer}, J.~L. 2018, \mnras, 476, 3661, \dodoi{10.1093/mnras/sty338}

\bibitem[{{Dom{\'\i}nguez S{\'a}nchez} {et~al.}(2022){Dom{\'\i}nguez S{\'a}nchez}, {Margalef}, {Bernardi}, \& {Huertas-Company}}]{DSanchez22}
{Dom{\'\i}nguez S{\'a}nchez}, H., {Margalef}, B., {Bernardi}, M., \& {Huertas-Company}, M. 2022, \mnras, 509, 4024, \dodoi{10.1093/mnras/stab3089}

\bibitem[{{Doubrawa} {et~al.}(2020){Doubrawa}, {Machado}, {Lagan{\'a}}, {Lima Neto}, {Monteiro-Oliveira}, \& {Cypriano}}]{Doubrawa20}
{Doubrawa}, L., {Machado}, R.~E.~G., {Lagan{\'a}}, T.~F., {et~al.} 2020, \mnras, 495, 2022, \dodoi{10.1093/mnras/staa1051}

\bibitem[{{Emsellem} {et~al.}(2007){Emsellem}, {Cappellari}, {Krajnovi{\'c}}, {van de Ven}, {Bacon}, {Bureau}, {Davies}, {de Zeeuw}, {Falc{\'o}n-Barroso}, {Kuntschner}, {McDermid}, {Peletier}, \& {Sarzi}}]{Emsellem07}
{Emsellem}, E., {Cappellari}, M., {Krajnovi{\'c}}, D., {et~al.} 2007, \mnras, 379, 401, \dodoi{10.1111/j.1365-2966.2007.11752.x}

\bibitem[{{Emsellem} {et~al.}(2011){Emsellem}, {Cappellari}, {Krajnovi{\'c}}, {Alatalo}, {Blitz}, {Bois}, {Bournaud}, {Bureau}, {Davies}, {Davis}, {de Zeeuw}, {Khochfar}, {Kuntschner}, {Lablanche}, {McDermid}, {Morganti}, {Naab}, {Oosterloo}, {Sarzi}, {Scott}, {Serra}, {van de Ven}, {Weijmans}, \& {Young}}]{Emsellem11}
---. 2011, \mnras, 414, 888, \dodoi{10.1111/j.1365-2966.2011.18496.x}

\bibitem[{Evert \& Baroni(2007)}]{zipfR}
Evert, S., \& Baroni, M. 2007, in Proceedings of the 45th Annual Meeting of the Association for Computational Linguistics, Posters and Demonstrations Sessions, Prague, Czech Republic, 29--32

\bibitem[{{Faber} {et~al.}(1997){Faber}, {Tremaine}, {Ajhar}, {Byun}, {Dressler}, {Gebhardt}, {Grillmair}, {Kormendy}, {Lauer}, \& {Richstone}}]{Faber97}
{Faber}, S.~M., {Tremaine}, S., {Ajhar}, E.~A., {et~al.} 1997, \aj, 114, 1771, \dodoi{10.1086/118606}

\bibitem[{{Ferguson} \& {Binggeli}(1994)}]{Ferguson94}
{Ferguson}, H.~C., \& {Binggeli}, B. 1994, \aapr, 6, 67, \dodoi{10.1007/BF01208252}

\bibitem[{{Fischer} {et~al.}(2019){Fischer}, {Dom{\'\i}nguez S{\'a}nchez}, \& {Bernardi}}]{Fischer19}
{Fischer}, J.~L., {Dom{\'\i}nguez S{\'a}nchez}, H., \& {Bernardi}, M. 2019, \mnras, 483, 2057, \dodoi{10.1093/mnras/sty3135}

\bibitem[{{Fogarty} {et~al.}(2015){Fogarty}, {Scott}, {Owers}, {Croom}, {Bekki}, {Houghton}, {van de Sande}, {D'Eugenio}, {Cecil}, {Colless}, {Bland-Hawthorn}, {Brough}, {Cortese}, {Davies}, {Jones}, {Pracy}, {Allen}, {Bryant}, {Goodwin}, {Green}, {Konstantopoulos}, {Lawrence}, {Lorente}, {Richards}, \& {Sharp}}]{Fogarty15}
{Fogarty}, L.~M.~R., {Scott}, N., {Owers}, M.~S., {et~al.} 2015, \mnras, 454, 2050, \dodoi{10.1093/mnras/stv2060}

\bibitem[{{Foreman-Mackey} {et~al.}(2013){Foreman-Mackey}, {Hogg}, {Lang}, \& {Goodman}}]{Foreman-Mackey13}
{Foreman-Mackey}, D., {Hogg}, D.~W., {Lang}, D., \& {Goodman}, J. 2013, \pasp, 125, 306, \dodoi{10.1086/670067}

\bibitem[{{Foster} {et~al.}(2017){Foster}, {van de Sande}, {D'Eugenio}, {Cortese}, {McDermid}, {Bland-Hawthorn}, {Brough}, {Bryant}, {Croom}, {Goodwin}, {Konstantopoulos}, {Lawrence}, {L{\'o}pez-S{\'a}nchez}, {Medling}, {Owers}, {Richards}, {Scott}, {Taranu}, {Tonini}, \& {Zafar}}]{Foster17}
{Foster}, C., {van de Sande}, J., {D'Eugenio}, F., {et~al.} 2017, \mnras, 472, 966, \dodoi{10.1093/mnras/stx1869}

\bibitem[{{Gallazzi} {et~al.}(2021){Gallazzi}, {Pasquali}, {Zibetti}, \& {Barbera}}]{Gallazzi21}
{Gallazzi}, A.~R., {Pasquali}, A., {Zibetti}, S., \& {Barbera}, F.~L. 2021, \mnras, 502, 4457, \dodoi{10.1093/mnras/stab265}

\bibitem[{{Gavazzi} {et~al.}(2005){Gavazzi}, {Donati}, {Cucciati}, {Sabatini}, {Boselli}, {Davies}, \& {Zibetti}}]{Gavazzi05}
{Gavazzi}, G., {Donati}, A., {Cucciati}, O., {et~al.} 2005, \aap, 430, 411, \dodoi{10.1051/0004-6361:20034571}

\bibitem[{{Graham} \& {Guzm{\'a}n}(2003)}]{Graham03}
{Graham}, A.~W., \& {Guzm{\'a}n}, R. 2003, \aj, 125, 2936, \dodoi{10.1086/374992}

\bibitem[{{Graham} {et~al.}(2018){Graham}, {Cappellari}, {Li}, {Mao}, {Bershady}, {Bizyaev}, {Brinkmann}, {Brownstein}, {Bundy}, {Drory}, {Law}, {Pan}, {Thomas}, {Wake}, {Weijmans}, {Westfall}, \& {Yan}}]{Graham18}
{Graham}, M.~T., {Cappellari}, M., {Li}, H., {et~al.} 2018, \mnras, 477, 4711, \dodoi{10.1093/mnras/sty504}

\bibitem[{{Hao} {et~al.}(2006){Hao}, {Mao}, {Deng}, {Xia}, \& {Wu}}]{Hao06}
{Hao}, C.~N., {Mao}, S., {Deng}, Z.~G., {Xia}, X.~Y., \& {Wu}, H. 2006, \mnras, 370, 1339, \dodoi{10.1111/j.1365-2966.2006.10545.x}

\bibitem[{{Hartigan} \& {Hartigan}(1985)}]{dip}
{Hartigan}, J.~A., \& {Hartigan}. 1985, "The Annals of Statistics", 13, 70, \dodoi{10.2307/2241-144:710954}

\bibitem[{{He} {et~al.}(2014){He}, {Hao}, \& {Xia}}]{He14}
{He}, Y.-Q., {Hao}, C.-N., \& {Xia}, X.-Y. 2014, Research in Astronomy and Astrophysics, 14, 144, \dodoi{10.1088/1674-4527/14/2/003}

\bibitem[{{Hern{\'a}ndez-Lang} {et~al.}(2022){Hern{\'a}ndez-Lang}, {Zenteno}, {Diaz-Ocampo}, {Cuevas}, {Clancy}, {Prado}, {Ald{\'a}s}, {Pallero}, {Monteiro-Oliveira}, {G{\'o}mez}, {Ramirez}, {Wynter}, {Carrasco}, {Hau}, {Stalder}, {McDonald}, {Bayliss}, {Floyd}, {Garmire}, {Katzenberger}, {Kim}, {Klein}, {Mahler}, {Nilo Castellon}, {Saro}, \& {Somboonpanyakul}}]{Hernandez-Lang22}
{Hern{\'a}ndez-Lang}, D., {Zenteno}, A., {Diaz-Ocampo}, A., {et~al.} 2022, \mnras, \dodoi{10.1093/mnras/stac2480}

\bibitem[{{Hopkins} {et~al.}(2009{\natexlab{a}}){Hopkins}, {Cox}, {Dutta}, {Hernquist}, {Kormendy}, \& {Lauer}}]{Hopkins09a}
{Hopkins}, P.~F., {Cox}, T.~J., {Dutta}, S.~N., {et~al.} 2009{\natexlab{a}}, \apjs, 181, 135, \dodoi{10.1088/0067-0049/181/1/135}

\bibitem[{{Hopkins} {et~al.}(2009{\natexlab{b}}){Hopkins}, {Lauer}, {Cox}, {Hernquist}, \& {Kormendy}}]{Hopkins09b}
{Hopkins}, P.~F., {Lauer}, T.~R., {Cox}, T.~J., {Hernquist}, L., \& {Kormendy}, J. 2009{\natexlab{b}}, \apjs, 181, 486, \dodoi{10.1088/0067-0049/181/2/486}

\bibitem[{{Jedrzejewski}(1987)}]{Jedrzejewski87}
{Jedrzejewski}, R.~I. 1987, \mnras, 226, 747, \dodoi{10.1093/mnras/226.4.747}

\bibitem[{{Kang} {et~al.}(2007){Kang}, {van den Bosch}, \& {Pasquali}}]{Kang07}
{Kang}, X., {van den Bosch}, F.~C., \& {Pasquali}, A. 2007, \mnras, 381, 389, \dodoi{10.1111/j.1365-2966.2007.12311.x}

\bibitem[{{Kass} \& {Raftery}(1995)}]{kass95}
{Kass}, R.~E., \& {Raftery}, A.~E. 1995, Journal of the American Statistical Association, 90, 773

\bibitem[{{Kelkar} {et~al.}(2020){Kelkar}, {Dwarakanath}, {Poggianti}, {Moretti}, {Monteiro-Oliveira}, {Machado}, {Lima-Neto}, {Fritz}, {Vulcani}, {Gullieuszik}, \& {Bettoni}}]{Kelkar20}
{Kelkar}, K., {Dwarakanath}, K.~S., {Poggianti}, B.~M., {et~al.} 2020, \mnras, 496, 442, \dodoi{10.1093/mnras/staa1547}

\bibitem[{{Kormendy}(1985)}]{Kormendy85}
{Kormendy}, J. 1985, \apjl, 292, L9, \dodoi{10.1086/184463}

\bibitem[{{Kormendy} \& {Bender}(1996)}]{Kormendy96}
{Kormendy}, J., \& {Bender}, R. 1996, \apjl, 464, L119, \dodoi{10.1086/310095}

\bibitem[{{Kormendy} \& {Bender}(2012)}]{Kormendy12}
---. 2012, \apjs, 198, 2, \dodoi{10.1088/0067-0049/198/1/2}

\bibitem[{{Kormendy} \& {Djorgovski}(1989)}]{Kormendy89}
{Kormendy}, J., \& {Djorgovski}, S. 1989, \araa, 27, 235, \dodoi{10.1146/annurev.aa.27.090189.001315}

\bibitem[{{Kormendy} {et~al.}(2009){Kormendy}, {Fisher}, {Cornell}, \& {Bender}}]{Kormendy09}
{Kormendy}, J., {Fisher}, D.~B., {Cornell}, M.~E., \& {Bender}, R. 2009, \apjs, 182, 216, \dodoi{10.1088/0067-0049/182/1/216}

\bibitem[{{Krajnovi{\'c}} {et~al.}(2013{\natexlab{a}}){Krajnovi{\'c}}, {Alatalo}, {Blitz}, {Bois}, {Bournaud}, {Bureau}, {Cappellari}, {Davies}, {Davis}, {de Zeeuw}, {Duc}, {Emsellem}, {Khochfar}, {Kuntschner}, {McDermid}, {Morganti}, {Naab}, {Oosterloo}, {Sarzi}, {Scott}, {Serra}, {Weijmans}, \& {Young}}]{Krajnovic13}
{Krajnovi{\'c}}, D., {Alatalo}, K., {Blitz}, L., {et~al.} 2013{\natexlab{a}}, \mnras, 432, 1768, \dodoi{10.1093/mnras/sts315}

\bibitem[{{Krajnovi{\'c}} {et~al.}(2013{\natexlab{b}}){Krajnovi{\'c}}, {Karick}, {Davies}, {Naab}, {Sarzi}, {Emsellem}, {Cappellari}, {Serra}, {de Zeeuw}, {Scott}, {McDermid}, {Weijmans}, {Davis}, {Alatalo}, {Blitz}, {Bois}, {Bureau}, {Bournaud}, {Crocker}, {Duc}, {Khochfar}, {Kuntschner}, {Morganti}, {Oosterloo}, \& {Young}}]{Krajnovic13b}
{Krajnovi{\'c}}, D., {Karick}, A.~M., {Davies}, R.~L., {et~al.} 2013{\natexlab{b}}, \mnras, 433, 2812, \dodoi{10.1093/mnras/stt905}

\bibitem[{{Krajnovi{\'c}} {et~al.}(2020){Krajnovi{\'c}}, {Ural}, {Kuntschner}, {Goudfrooij}, {Wolfe}, {Cappellari}, {Davies}, {de Zeeuw}, {Duc}, {Emsellem}, {Karick}, {McDermid}, {Mei}, \& {Naab}}]{Krajnovic20}
{Krajnovi{\'c}}, D., {Ural}, U., {Kuntschner}, H., {et~al.} 2020, \aap, 635, A129, \dodoi{10.1051/0004-6361/201937040}

\bibitem[{{Kroupa}(2001)}]{Kroupa01}
{Kroupa}, P. 2001, \mnras, 322, 231, \dodoi{10.1046/j.1365-8711.2001.04022.x}

\bibitem[{{Lauer}(1985)}]{Lauer85b}
{Lauer}, T.~R. 1985, \mnras, 216, 429, \dodoi{10.1093/mnras/216.2.429}

\bibitem[{{Lauer}(2012)}]{Lauer12}
---. 2012, \apj, 759, 64, \dodoi{10.1088/0004-637X/759/1/64}

\bibitem[{{Law} {et~al.}(2015){Law}, {Yan}, {Bershady}, {Bundy}, {Cherinka}, {Drory}, {MacDonald}, {S{\'a}nchez-Gallego}, {Wake}, {Weijmans}, {Blanton}, {Klaene}, {Moran}, {Sanchez}, \& {Zhang}}]{Law15}
{Law}, D.~R., {Yan}, R., {Bershady}, M.~A., {et~al.} 2015, \aj, 150, 19, \dodoi{10.1088/0004-6256/150/1/19}

\bibitem[{{Law} {et~al.}(2016){Law}, {Cherinka}, {Yan}, {Andrews}, {Bershady}, {Bizyaev}, {Blanc}, {Blanton}, {Bolton}, {Brownstein}, {Bundy}, {Chen}, {Drory}, {D'Souza}, {Fu}, {Jones}, {Kauffmann}, {MacDonald}, {Masters}, {Newman}, {Parejko}, {S{\'a}nchez-Gallego}, {S{\'a}nchez}, {Schlegel}, {Thomas}, {Wake}, {Weijmans}, {Westfall}, \& {Zhang}}]{Law16}
{Law}, D.~R., {Cherinka}, B., {Yan}, R., {et~al.} 2016, \aj, 152, 83, \dodoi{10.3847/0004-6256/152/4/83}

\bibitem[{{Lin} {et~al.}(2018){Lin}, {Huang}, \& {Chen}}]{Lin18}
{Lin}, Y.-T., {Huang}, H.-J., \& {Chen}, Y.-C. 2018, \aj, 155, 188, \dodoi{10.3847/1538-3881/aab5b4}

\bibitem[{{Lin} \& {Mohr}(2004)}]{Lin04}
{Lin}, Y.-T., \& {Mohr}, J.~J. 2004, \apj, 617, 879, \dodoi{10.1086/425412}

\bibitem[{{Lin} \& {Mohr}(2007)}]{Lin07}
---. 2007, \apjs, 170, 71, \dodoi{10.1086/513565}

\bibitem[{{Lin} {et~al.}(2010){Lin}, {Shen}, {Strauss}, {Richards}, \& {Lunnan}}]{Lin10}
{Lin}, Y.-T., {Shen}, Y., {Strauss}, M.~A., {Richards}, G.~T., \& {Lunnan}, R. 2010, \apj, 723, 1119, \dodoi{10.1088/0004-637X/723/2/1119}

\bibitem[{{Lintott} {et~al.}(2008){Lintott}, {Schawinski}, {Slosar}, {Land}, {Bamford}, {Thomas}, {Raddick}, {Nichol}, {Szalay}, {Andreescu}, {Murray}, \& {Vandenberg}}]{Lintott08}
{Lintott}, C.~J., {Schawinski}, K., {Slosar}, A., {et~al.} 2008, \mnras, 389, 1179, \dodoi{10.1111/j.1365-2966.2008.13689.x}

\bibitem[{{Machado} {et~al.}(2015){Machado}, {Monteiro-Oliveira}, {Lima Neto}, \& {Cypriano}}]{Machado15b}
{Machado}, R.~E.~G., {Monteiro-Oliveira}, R., {Lima Neto}, G.~B., \& {Cypriano}, E.~S. 2015, \mnras, 451, 3309, \dodoi{10.1093/mnras/stv1162}

\bibitem[{{Machado} {et~al.}(2024){Machado}, {Volert}, {Albuquerque}, {Monteiro-Oliveira}, \& {Lima Neto}}]{Machado24}
{Machado}, R. E.~G., {Volert}, R.~C., {Albuquerque}, R.~P., {Monteiro-Oliveira}, R., \& {Lima Neto}, G.~B. 2024, \apj, 970, 160, \dodoi{10.3847/1538-4357/ad5350}

\bibitem[{Martin {et~al.}(2011)Martin, Quinn, \& Park}]{MCMCpack}
Martin, A.~D., Quinn, K.~M., \& Park, J.~H. 2011, Journal of Statistical Software, 42, 22.
\newblock \url{http://www.jstatsoft.org/v42/i09/}

\bibitem[{{McDermid} {et~al.}(2015){McDermid}, {Alatalo}, {Blitz}, {Bournaud}, {Bureau}, {Cappellari}, {Crocker}, {Davies}, {Davis}, {de Zeeuw}, {Duc}, {Emsellem}, {Khochfar}, {Krajnovi{\'c}}, {Kuntschner}, {Morganti}, {Naab}, {Oosterloo}, {Sarzi}, {Scott}, {Serra}, {Weijmans}, \& {Young}}]{McDermid15}
{McDermid}, R.~M., {Alatalo}, K., {Blitz}, L., {et~al.} 2015, \mnras, 448, 3484, \dodoi{10.1093/mnras/stv105}

\bibitem[{{Milvang-Jensen} \& {J{\o}rgensen}(1999)}]{Milvang-Jensen99}
{Milvang-Jensen}, B., \& {J{\o}rgensen}, I. 1999, Baltic Astronomy, 8, 535, \dodoi{10.1515/astro-1999-0408}

\bibitem[{{Mitsuda} {et~al.}(2017){Mitsuda}, {Doi}, {Morokuma}, {Suzuki}, {Yasuda}, {Perlmutter}, {Aldering}, \& {Meyers}}]{Mitsuda17}
{Mitsuda}, K., {Doi}, M., {Morokuma}, T., {et~al.} 2017, \apj, 834, 109, \dodoi{10.3847/1538-4357/834/2/109}

\bibitem[{{Monteiro-Oliveira}(2022)}]{Monteiro-Oliveira22b}
{Monteiro-Oliveira}, R. 2022, \mnras, 515, 3674, \dodoi{10.1093/mnras/stac2053}

\bibitem[{{Monteiro-Oliveira} {et~al.}(2017{\natexlab{a}}){Monteiro-Oliveira}, {Cypriano}, {Machado}, {Lima Neto}, {Ribeiro}, {Sodr{\'e}}, \& {Dupke}}]{Monteiro-Oliveira17a}
{Monteiro-Oliveira}, R., {Cypriano}, E.~S., {Machado}, R.~E.~G., {et~al.} 2017{\natexlab{a}}, \mnras, 466, 2614, \dodoi{10.1093/mnras/stw3238}

\bibitem[{{Monteiro-Oliveira} {et~al.}(2018){Monteiro-Oliveira}, {Cypriano}, {Vitorelli}, {Ribeiro}, {Sodr{\'e}}, {Dupke}, \& {Mendes de Oliveira}}]{Monteiro-Oliveira18}
{Monteiro-Oliveira}, R., {Cypriano}, E.~S., {Vitorelli}, A.~Z., {et~al.} 2018, \mnras, 481, 1097, \dodoi{10.1093/mnras/sty2349}

\bibitem[{{Monteiro-Oliveira} {et~al.}(2020){Monteiro-Oliveira}, {Doubrawa}, {Machado}, {Lima Neto}, {Castejon}, \& {Cypriano}}]{Monteiro-Oliveira20}
{Monteiro-Oliveira}, R., {Doubrawa}, L., {Machado}, R.~E.~G., {et~al.} 2020, \mnras, 495, 2007, \dodoi{10.1093/mnras/staa1218}

\bibitem[{{Monteiro-Oliveira} {et~al.}(2017{\natexlab{b}}){Monteiro-Oliveira}, {Lima Neto}, {Cypriano}, {Machado}, {Capelato}, {Lagan{\'a}}, {Durret}, \& {Bagchi}}]{Monteiro-Oliveira17b}
{Monteiro-Oliveira}, R., {Lima Neto}, G.~B., {Cypriano}, E.~S., {et~al.} 2017{\natexlab{b}}, \mnras, 468, 4566, \dodoi{10.1093/mnras/stx791}

\bibitem[{{Monteiro-Oliveira} {et~al.}(2022){Monteiro-Oliveira}, {Morell}, {Sampaio}, {Ribeiro}, \& {de Carvalho}}]{Monteiro-Oliveira22a}
{Monteiro-Oliveira}, R., {Morell}, D.~F., {Sampaio}, V.~M., {Ribeiro}, A.~L.~B., \& {de Carvalho}, R.~R. 2022, \mnras, 509, 3470, \dodoi{10.1093/mnras/stab3225}

\bibitem[{{Monteiro-Oliveira} {et~al.}(2021){Monteiro-Oliveira}, {Soja}, {Ribeiro}, {Bagchi}, {Sankhyayan}, {Candido}, \& {Flores}}]{Monteiro-Oliveira21}
{Monteiro-Oliveira}, R., {Soja}, A.~C., {Ribeiro}, A.~L.~B., {et~al.} 2021, \mnras, 501, 756, \dodoi{10.1093/mnras/staa3575}

\bibitem[{{Moura} {et~al.}(2021){Moura}, {Machado}, \& {Monteiro-Oliveira}}]{Moura21}
{Moura}, M.~T., {Machado}, R. E.~G., \& {Monteiro-Oliveira}, R. 2021, \mnras, 500, 1858, \dodoi{10.1093/mnras/staa3399}

\bibitem[{{Naab} {et~al.}(1999){Naab}, {Burkert}, \& {Hernquist}}]{Naab99}
{Naab}, T., {Burkert}, A., \& {Hernquist}, L. 1999, \apjl, 523, L133, \dodoi{10.1086/312275}

\bibitem[{{Naab} {et~al.}(2006){Naab}, {Khochfar}, \& {Burkert}}]{Naab06}
{Naab}, T., {Khochfar}, S., \& {Burkert}, A. 2006, \apjl, 636, L81, \dodoi{10.1086/500205}

\bibitem[{{Naab} {et~al.}(2014){Naab}, {Oser}, {Emsellem}, {Cappellari}, {Krajnovi{\'c}}, {McDermid}, {Alatalo}, {Bayet}, {Blitz}, {Bois}, {Bournaud}, {Bureau}, {Crocker}, {Davies}, {Davis}, {de Zeeuw}, {Duc}, {Hirschmann}, {Johansson}, {Khochfar}, {Kuntschner}, {Morganti}, {Oosterloo}, {Sarzi}, {Scott}, {Serra}, {van de Ven}, {Weijmans}, \& {Young}}]{Naab14}
{Naab}, T., {Oser}, L., {Emsellem}, E., {et~al.} 2014, \mnras, 444, 3357, \dodoi{10.1093/mnras/stt1919}

\bibitem[{{Nair} \& {Abraham}(2010)}]{Nair10}
{Nair}, P.~B., \& {Abraham}, R.~G. 2010, \apjs, 186, 427, \dodoi{10.1088/0067-0049/186/2/427}

\bibitem[{{Nieto} \& {Bender}(1989)}]{Nieto89}
{Nieto}, J.~L., \& {Bender}, R. 1989, \aap, 215, 266

\bibitem[{{Pandge} {et~al.}(2019){Pandge}, {Monteiro-Oliveira}, {Bagchi}, {Simionescu}, {Limousin}, \& {Raychaudhury}}]{Pandge19}
{Pandge}, M.~B., {Monteiro-Oliveira}, R., {Bagchi}, J., {et~al.} 2019, \mnras, 482, 5093, \dodoi{10.1093/mnras/sty2937}

\bibitem[{{Pasquali} {et~al.}(2007){Pasquali}, {van den Bosch}, \& {Rix}}]{Pasquali07}
{Pasquali}, A., {van den Bosch}, F.~C., \& {Rix}, H.~W. 2007, \apj, 664, 738, \dodoi{10.1086/518856}

\bibitem[{{Pellegrini}(1999)}]{Pellegrini99}
{Pellegrini}, S. 1999, \aap, 351, 487, \dodoi{10.48550/arXiv.astro-ph/9909458}

\bibitem[{{Penoyre} {et~al.}(2017){Penoyre}, {Moster}, {Sijacki}, \& {Genel}}]{Penoyre17}
{Penoyre}, Z., {Moster}, B.~P., {Sijacki}, D., \& {Genel}, S. 2017, \mnras, 468, 3883, \dodoi{10.1093/mnras/stx762}

\bibitem[{{R Core Team}(2014)}]{R}
{R Core Team}. 2014, R: A Language and Environment for Statistical Computing, R Foundation for Statistical Computing, Vienna, Austria.
\newblock \url{http://www.R-project.org/}

\bibitem[{{S{\'a}nchez-Bl{\'a}zquez} {et~al.}(2006){S{\'a}nchez-Bl{\'a}zquez}, {Peletier}, {Jim{\'e}nez-Vicente}, {Cardiel}, {Cenarro}, {Falc{\'o}n-Barroso}, {Gorgas}, {Selam}, \& {Vazdekis}}]{Sanchez-Blazquez06}
{S{\'a}nchez-Bl{\'a}zquez}, P., {Peletier}, R.~F., {Jim{\'e}nez-Vicente}, J., {et~al.} 2006, \mnras, 371, 703, \dodoi{10.1111/j.1365-2966.2006.10699.x}

\bibitem[{{Sarzi} {et~al.}(2013){Sarzi}, {Alatalo}, {Blitz}, {Bois}, {Bournaud}, {Bureau}, {Cappellari}, {Crocker}, {Davies}, {Davis}, {de Zeeuw}, {Duc}, {Emsellem}, {Khochfar}, {Krajnovi{\'c}}, {Kuntschner}, {Lablanche}, {McDermid}, {Morganti}, {Naab}, {Oosterloo}, {Scott}, {Serra}, {Young}, \& {Weijmans}}]{Sarzi13}
{Sarzi}, M., {Alatalo}, K., {Blitz}, L., {et~al.} 2013, \mnras, 432, 1845, \dodoi{10.1093/mnras/stt062}

\bibitem[{{Smee} {et~al.}(2013){Smee}, {Gunn}, {Uomoto}, {Roe}, {Schlegel}, {Rockosi}, {Carr}, {Leger}, {Dawson}, {Olmstead}, {Brinkmann}, {Owen}, {Barkhouser}, {Honscheid}, {Harding}, {Long}, {Lupton}, {Loomis}, {Anderson}, {Annis}, {Bernardi}, {Bhardwaj}, {Bizyaev}, {Bolton}, {Brewington}, {Briggs}, {Burles}, {Burns}, {Castander}, {Connolly}, {Davenport}, {Ebelke}, {Epps}, {Feldman}, {Friedman}, {Frieman}, {Heckman}, {Hull}, {Knapp}, {Lawrence}, {Loveday}, {Mannery}, {Malanushenko}, {Malanushenko}, {Merrelli}, {Muna}, {Newman}, {Nichol}, {Oravetz}, {Pan}, {Pope}, {Ricketts}, {Shelden}, {Sandford}, {Siegmund}, {Simmons}, {Smith}, {Snedden}, {Schneider}, {SubbaRao}, {Tremonti}, {Waddell}, \& {York}}]{Smee13}
{Smee}, S.~A., {Gunn}, J.~E., {Uomoto}, A., {et~al.} 2013, \aj, 146, 32, \dodoi{10.1088/0004-6256/146/2/32}

\bibitem[{{Soja} {et~al.}(2018){Soja}, {Sodr{\'e}}, {Monteiro-Oliveira}, {Cypriano}, \& {Lima Neto}}]{Soja18}
{Soja}, A.~C., {Sodr{\'e}}, L., {Monteiro-Oliveira}, R., {Cypriano}, E.~S., \& {Lima Neto}, G.~B. 2018, \mnras, 477, 3279, \dodoi{10.1093/mnras/sty638}

\bibitem[{Therneau(2018)}]{deming}
Therneau, T. 2018, deming: Deming, Theil-Sen, Passing-Bablock and Total Least Squares Regression.
\newblock \url{https://CRAN.R-project.org/package=deming}

\bibitem[{{Thomas} {et~al.}(2005){Thomas}, {Maraston}, {Bender}, \& {Mendes de Oliveira}}]{Thomas05}
{Thomas}, D., {Maraston}, C., {Bender}, R., \& {Mendes de Oliveira}, C. 2005, \apj, 621, 673, \dodoi{10.1086/426932}

\bibitem[{{Tremblay} \& {Merritt}(1996)}]{Tremblay96}
{Tremblay}, B., \& {Merritt}, D. 1996, \aj, 111, 2243, \dodoi{10.1086/117959}

\bibitem[{{Villaume} {et~al.}(2017){Villaume}, {Conroy}, {Johnson}, {Rayner}, {Mann}, \& {van Dokkum}}]{Villaume17}
{Villaume}, A., {Conroy}, C., {Johnson}, B., {et~al.} 2017, \apjs, 230, 23, \dodoi{10.3847/1538-4365/aa72ed}

\bibitem[{{Wall} \& {Jenkins}(2012)}]{Wall12}
{Wall}, J.~V., \& {Jenkins}, C.~R. 2012, {Practical Statistics for Astronomers}

\bibitem[{{Wand}(2015)}]{KernSmooth}
{Wand}, M. 2015, Functions for Kernel Smoothing.
\newblock \url{https://CRAN.R-project.org/package=KernSmooth}

\bibitem[{{Watson} {et~al.}(2022){Watson}, {Davies}, {van de Sande}, {Brough}, {Croom}, {D'Eugenio}, {Glazebrook}, {Groves}, {L{\'o}pez-S{\'a}nchez}, {Scott}, {Vaughan}, {Walcher}, {Bland-Hawthorn}, {Bryant}, {Goodwin}, {Lawrence}, {Lorente}, {Owers}, \& {Richards}}]{Watson22}
{Watson}, P.~J., {Davies}, R.~L., {van de Sande}, J., {et~al.} 2022, \mnras, 513, 5076, \dodoi{10.1093/mnras/stac1221}

\bibitem[{{Weijmans} {et~al.}(2014){Weijmans}, {de Zeeuw}, {Emsellem}, {Krajnovi{\'c}}, {Lablanche}, {Alatalo}, {Blitz}, {Bois}, {Bournaud}, {Bureau}, {Cappellari}, {Crocker}, {Davies}, {Davis}, {Duc}, {Khochfar}, {Kuntschner}, {McDermid}, {Morganti}, {Naab}, {Oosterloo}, {Sarzi}, {Scott}, {Serra}, {Verdoes Kleijn}, \& {Young}}]{Weijmans14}
{Weijmans}, A.-M., {de Zeeuw}, P.~T., {Emsellem}, E., {et~al.} 2014, \mnras, 444, 3340, \dodoi{10.1093/mnras/stu1603}

\bibitem[{{Willett} {et~al.}(2013){Willett}, {Lintott}, {Bamford}, {Masters}, {Simmons}, {Casteels}, {Edmondson}, {Fortson}, {Kaviraj}, {Keel}, {Melvin}, {Nichol}, {Raddick}, {Schawinski}, {Simpson}, {Skibba}, {Smith}, \& {Thomas}}]{Willett13}
{Willett}, K.~W., {Lintott}, C.~J., {Bamford}, S.~P., {et~al.} 2013, \mnras, 435, 2835, \dodoi{10.1093/mnras/stt1458}

\bibitem[{{Zenteno} {et~al.}(2025){Zenteno}, {Kluge}, {Kharkrang}, {Hernandez-Lang}, {Damke}, {Saro}, {Monteiro-Oliveira}, {Carrasco}, {Salvato}, {Comparat}, {Fabricius}, {Snigula}, {Arevalo}, {Cuevas}, {Nilo Castellon}, {Ramirez}, {V\textbackslash'eliz Astudillo}, {Landriau}, {Myers}, {Schlafly}, {Valdes}, {Weaver}, {Mohr}, {Grandis}, {Klein}, {Liu}, {Bulbul}, {Zhang}, {Sanders}, {Bahar}, {Ghirardini}, {Ramos}, \& {Balzer}}]{Zenteno25}
{Zenteno}, A., {Kluge}, M., {Kharkrang}, R., {et~al.} 2025, arXiv e-prints, arXiv:2503.21066, \dodoi{10.48550/arXiv.2503.21066}

\bibitem[{{Zheng} {et~al.}(2023){Zheng}, {R{\"o}ttgering}, {van der Wel}, \& {Cappellari}}]{Zheng23}
{Zheng}, X., {R{\"o}ttgering}, H., {van der Wel}, A., \& {Cappellari}, M. 2023, \aap, 673, A12, \dodoi{10.1051/0004-6361/202245405}

\bibitem[{{Zhu} {et~al.}(2010){Zhu}, {Blanton}, \& {Moustakas}}]{Zhu10}
{Zhu}, G., {Blanton}, M.~R., \& {Moustakas}, J. 2010, \apj, 722, 491, \dodoi{10.1088/0004-637X/722/1/491}

\bibitem[{{Zhu} {et~al.}(2023){Zhu}, {Lu}, {Cappellari}, {Li}, {Mao}, \& {Gao}}]{Zhu23}
{Zhu}, K., {Lu}, S., {Cappellari}, M., {et~al.} 2023, \mnras, 522, 6326, \dodoi{10.1093/mnras/stad1299}

\end{thebibliography}
\bibliographystyle{aasjournal}

\end{document}